\documentclass{aa}
\usepackage{txfonts}
\usepackage[dvips]{graphicx}
\newcommand{\mcu}[1]{\multicolumn{1}{c}{#1}}

\def\Msun{\mbox{$M_\odot$}}
\def\Lsun{\mbox{$L_\odot$}}

\def\CSI{C$^{34}$S}
\def\CO{$^{12}$CO}
\def\COI{\mbox{$^{13}$CO}}

\def\COIII{\mbox{C$^{17}$O}}

\def\MCN{\mbox{CH$_3$CN}}
\def\MCNI{$^{13}$CH$_3$CN}
\def\MCNII{C${{\rm H}_3}^{13}$CN}
\def\HM{H$_2$}
\def\HII{H{\sc ii}}

\def\kms{\mbox{km~s$^{-1}$}}
\def\cmc{cm$^{-3}$}

\begin{document}
\title{
Dissecting a hot molecular core: The case of G31.41+0.31\thanks{Based on
observations carried out with the Submillimeter Array. The Submillimeter Array
is a joint project between the Smithsonian Astrophysical Observatory and the
Academia Sinica Institute of Astronomy and Astrophysics and is funded by the
Smithsonian Institution and the Academia Sinica.}
}
\author{R. Cesaroni \inst{1} \and M.T. Beltr\'an \inst{1} \and Q. Zhang \inst{2}
	\and H. Beuther \inst{3} \and C. Fallscheer \inst{3,4}
	}
\institute{
	   Osservatorio Astrofisico di Arcetri, INAF, Largo E. Fermi 5,
           I-50125 Firenze, Italy \email{cesa@arcetri.astro.it}
\and
	   Harvard-Smithsonian Center for Astrophysics, Cambridge, MA 02138,
	   U.S.A. \email{qzhang@cfa.harvard.edu}
\and
	   Max-Planck-Institut f\"ur Astronomie, K\"onigstuhl 17,
	   D-69117 Heidelberg, Germany \email{beuther@mpia-hd.mpg.de}
\and
	   Department of Physics and Astronomy, University of 
	   Victoria, 3800 Finnerty Road, Victoria, BC V8P 5C2, Canada
	   \email{cassandra.fallscheer@nrc-cnrc.gc.ca}
}
\offprints{R. Cesaroni, \email{cesa@arcetri.astro.it}}
\date{Received date; accepted date}

\titlerunning{}
\authorrunning{}

\abstract{
The role of disks in the formation of high-mass stars is still a matter of debate
but the detection of circumstellar disks around O-type stars would have a
profound impact on high-mass star formation theories.
}{
We made a detailed observational analysis of a well known hot molecular
core lying in the high-mass star-forming region G31.41+0.31. This core
is believed to contain deeply embedded massive stars and
presents a velocity gradient that has been interpreted either as rotation or
as expansion, depending on the authors. Our aim was to shed light on this question
and possibly prepare the ground for higher resolution ALMA observations which
could directly detect circumstellar disks around the embedded massive stars.
}{
Observations at sub-arcsecond resolution were performed with the Submillimeter
Array in methyl cyanide, a typical hot molecular core tracer, and \CO\ and
\COI, well known outflow tracers. We also obtained sensitive continuum maps
at 1.3~mm.
}{
Our findings confirm the existence of a sharp velocity gradient across the
core, but cannot confirm the existence of a bipolar outflow perpendicular
to it. The improved angular resolution and sampling of the uv plane allow us
to attain higher quality channel maps of the \MCN\ lines with respect to
previous studies and thus significantly improve our knowledge of the
structure and kinematics of the hot molecular core.
}{
While no conclusive argument can rule out any of the two interpretations
(rotation or expansion) proposed
to explain the velocity gradient observed in the core, in our opinion the
observational evidence collected so far indicates the rotating toroid as the
most likely scenario. The outflow hypothesis appears less plausible,
because the dynamical time scale is too short compared to that needed to
form species such as \MCN, and the mass loss
and momentum rates estimated from our measurements appear too high.
}

\maketitle

\section{Introduction}
\label{sint}

High-mass stars are usually defined as those exceeding $\sim$8~\Msun, based on
the fact that stars above this mass limit do not have a pre-main sequence phase
(Palla \& Stahler \cite{past}). This means that accretion is ongoing until
the star ignites hydrogen burning and reaches the zero-age main sequence (ZAMS).
At this point the strong radiation pressure may halt and even reverse infall
and thus stop further growth of the stellar mass.
This led to the so-called ``radiation pressure problem''. Recent studies have
demonstrated that this limitation holds only in spherical symmetry.
As first envisioned by Nakano~(\cite{nakano}) and recently demonstrated by
Krumholz et al.~(\cite{krum}) and Kuiper et al.~(\cite{kuip}), accretion
through a circumstellar disk can explain the formation of stars up to the
upper limit of the initial mass function, by allowing part of the photons to
escape along the disk axis and boosting the ram pressure of the accreting gas
through the small disk solid angle. It also appears that the powerful
ionizing fluxes from these OB-type stars are not sufficient to destroy the
disk, which eventually turns into an ionized, rotating accretion flow close
to the star (Sollins et al.~\cite{sollins}; Keto \cite{keto07}).

For these reasons it seems established that circumstellar accretion disks
play a crucial role in the formation of {\it all} stars and not only
solar-type stars. This theoretical result contrasts with
the limited observational evidence of disks in high-mass (proto)stars.
Only in recent years the number of disk candidates associated with luminous
young stellar objects (YSOs) has significantly increased, mostly owing to the
improvement of (sub)millimeter interferometers in terms of angular resolution
and sensitivity. The main problems that one has to face in this type of search
are the large distances to the sources (typically a few kpc) and the confusion
caused by stellar crowding (OB-type stars form in clusters). These factors
may explain
the failure to detect disks in O-type stars, as opposed to the number of
detections obtained for B-type stars (Cesaroni et al.~\cite{ppv}).
In association with the most luminous YSOs one finds only huge ($\la$0.1~pc),
massive (a few 100~\Msun) cores, with velocity gradients suggesting rotation.
These objects, named ``toroids'', are likely non-equilibrium structures,
because the ratio between the accretion time scale and the rotation period is
shorter than for disks: this implies that the toroid does not have enough
time to adjust its structure to the new fresh material falling onto it
(Cesaroni et al.~\cite{ppv}; Beltr\'an et al.~\cite{bel11}).

With this in mind, one can see that understanding the formation of the
most massive stars ($>20~\Msun$) may benefit from a detailed investigation
of toroids, also because these objects might be hosting
true circumstellar disks in their interiors. Moreover, their mere existence
may set tighter constraints on theoretical models. Despite the
number of candidates, the existence of rotating toroids is still a matter
of debate. What is questioned is the nature of the velocity gradient,
which is sometimes interpreted as expansion instead of rotation (see
e.g. Gibb et al.~\cite{gibb} and Araya et al.~\cite{araya} and references
therein), thus
suggesting that one might be seeing a compact bipolar outflow rather than
a rotating core. In an attempt to distinguish between these two possibilities
and more in general to shed light on these intriguing objects, we have
focused our attention on one of the best examples: the hot molecular core
(HMC) G31.41+0.31 (hereafter G31.41).

This prototypical HMC is located at a kinematic distance of 7.9~kpc and was
originally imaged in the high-excitation (4,4) inversion transition of
ammonia (Cesaroni et al.~\cite{cesa94}) and in the (6--5) rotational
transitions of methyl cyanide (\MCN; Cesaroni et al.~\cite{cesag31}). The latter showed the
existence of a striking velocity gradient (centered at an LSR velocity of
$\sim$96.5~\kms) across the core in the NE--SW
direction, already suggested by the distribution and velocities of OH masers
(Gaume \& Mutel~\cite{gamu}). Follow-up interferometric observations with
better angular resolution and in high-energy tracers have confirmed this
result and revealed the presence of deeply embedded YSOs, which in all
likelihood explains the temperature increase toward the core center
(Beltr\'an et al.~\cite{bel04}, \cite{bel05}, hereafter BEL04 and BEL05;
Cesaroni et al.~\cite{cesa10}).
The G31.41 HMC is separated by $\sim$5\arcsec\ from an ultracompact (UC)
\HII\ region, and overlaps in projection on a diffuse halo of free-free
emission, possibly associated with the UC~\HII\ region itself (see e.g.
Fig.~2c of Cesaroni et al.~\cite{cesa98}). The SPITZER/GLIMPSE images
(Benjamin et al.~\cite{benj})
show
that the HMC lies in a complex pc-scale region where both extended emission
and multiple stellar sources are detected. All these facts complicate the
interpretation of the HMC structure and its relationship with the molecular
surroundings and call for additional high-quality observations.

An important aspect of a proper interpretation of the velocity gradient
observed in the HMC is the existence of an associated molecular outflow.
A reasonable assumption, suggested by the analogy with low-mass YSOs,
is that a disk actively undergoing accretion must be associated with a
bipolar jet/outflow expanding along the disk rotation axis. Therefore,
a way to distinguish between the two interpretations of the HMC
velocity gradient (expansion vs rotation) is to search for a molecular
outflow perpendicular to the gradient. Previous interferometric observations
in the \COI(1--0) line have suggested the presence of a collimated outflow
directed SE--NW (Olmi et al.~\cite{olmi96b}), but the poor uv coverage,
especially on the shortest baselines, makes this result questionable.
We have therefore conducted new
interferometric observations at 1.3~mm with the Submillimeter Array (SMA) in
the typical HMC tracer \MCN\ and in standard outflow tracers such as \CO\ and
\COI. In order to be sensitive to extended structures filtered out by
the interferometer, we have also mapped the region over $\sim$2\arcmin\
with the IRAM 30-m telescope. The observational details are given in
Sect.~\ref{sobs}, while the results are illustrated in Sect.~\ref{sres}
and discussed in Sect.~\ref{sdis}. Finally, the conclusions are drawn in
Sect.~\ref{scon}.

\section{Observations and data reduction}
\label{sobs}

Our observations have been performed with the SMA interferometer and
IRAM 30-m telescope. The former was used with the aim to image the \CO\ and
\COI\ (2--1) lines and the \MCN(12--11) transitions, but
the 2~GHz bandwidth in both the LSB and USB allowed us to cover a much
larger number of lines from a plethora of different molecules. Despite the
richness of the spectrum, in this article we will present only the results
obtained for the above mentioned lines, which are the best tracers for our
purposes. The 30-m telescope was used to fill the zero-spacing of the
SMA maps in the \CO\ and \COI\ lines and thus help establishing the
morphology of the emission at all velocities and on all scales.
Technical details of the observations are given in the next
two sections.

\begin{table}
\begin{flushleft}
\caption[]{Parameters of SMA images. The RMS of the lines is estimated in
each channel}
\label{tobs}
\begin{tabular}{cccc}
\hline
 \mcu{Tracer} & Resolution & \mcu{1\,$\sigma$ RMS} &    synth. beam, PA \\
 & (\kms) & \mcu{(mJy/beam)} &      \\
\hline
 cont. & --- &    ~~7 &     0\farcs88 x 0\farcs72, 56\fdg6 \\
 \MCN\ & 0.6 &    90 &     0\farcs89 x 0\farcs75, 52\fdg8 \\
 \MCNI\ & 0.6 &    90 &     0\farcs89 x 0\farcs75, 52\fdg8 \\
 \CO\  & 1.0 &    60 &     0\farcs89 x 0\farcs73, 53\fdg4 \\
 \COI\ & 1.0 &    50 &     0\farcs88 x 0\farcs75, 52\fdg7 \\
\hline
\end{tabular}
\end{flushleft}
\end{table}

\subsection{SMA interferometer}

Observations of G31.41 were carried out with the SMA \footnote{The Submillimeter
Array is a joint project between the Smithsonian Astrophysical Observatory
and the Academia Sinica Institute of Astronomy and Astrophysics, and is
funded by the Smithsonian Institution and the Academia Sinica.} (Ho et al.
\cite{ho2004}) in the 230~GHz band in the compact and extended
configurations. The correlator was configured to a uniform spectral
resolution of 0.41~MHz (0.6~\kms) over the entire 4 GHz spectral window. With IF
frequencies of 4 to 6 GHz, the observations covered the rest frequencies from
219.3--221.3 GHz in the lower side band (LSB), and 229.3--231.3 GHz in the
upper side band (USB). The compact configuration data were obtained on 2007
July 09 with eight antennas. With 225~GHz zenith opacity of 0.05 to 0.1, the
typical double side-band system temperatures were around 200~K. The very
extended data were obtained on 2007 May 21 with seven antennas. At 225~GHz the
zenith opacity was about 0.2 and the double side-band system temperatures
were around 400~K. For both tracks, we used Vesta and 3C\,273 for flux and
bandpass calibrations. The time-dependent gains were calibrated using
1751+096 and 1830+063. The phase center of the observations was
$\alpha$(J2000)=18$^{\rm h}$47$^{\rm m}$34\fs315,
$\delta$(J2000)=--01\degr12\arcmin45\farcs9. The primary beam of the 6-m
antennas is about 55\arcsec\ at the operating frequencies.

The visibility data were calibrated in the IDL superset MIR\footnote{
The MIR cookbook can be found at
https://www.cfa.harvard.edu/$\sim$cqi/mircook.html
}
and MIRIAD (Sault et al. \cite{sault}), and were exported to MIRIAD format
for imaging. The projected baselines of the combined visibilities from both
configurations range from 12~m to 500~m. The continuum data were constructed
from line free channels and the continuum was subtracted from the line uv data.

Continuum and channel maps of the \MCN\ and \MCNII\ lines were created and
cleaned in MIRIAD by weighting the data with ``robust=0'' to find a
compromise between angular resolution and sensitivity to extended
structures. The resulting synthesized beam is approximately
$0\farcs88\times0\farcs75$, with P.A. 53\degr. The $1\,\sigma$ rms
is 5.8~mJy/beam in the continuum, and 89~mJy/beam in the line images with
0.6~\kms\ spectral resolution.
The \CO\ and \COI\ (2--1) line data were imaged in a similar fashion, but
using natural weighting to enhance extended emission. The resulting
synthesized beam is $0\farcs89\times1\farcs73$ with P.A. 53\degr. The
$1\,\sigma$ rms in the 1~\kms\ channel maps is 50~mJy/beam.

The \CO\ and \COI\ data were also combined with the 30-m data (see
Sect.~\ref{ssd}) to recover extended structure resolved out by the
interferometer. The single-dish maps were Fourier-transformed and then
suitably sampled in the uv domain, then merged with the SMA data.
Finally natural weighted maps were created using the same procedure
as for the non-merged data. The resulting synthesized beam is
$2\farcs3\times1\farcs3$ with P.A. 63\degr and the $1\,\sigma$ rms in the
1~\kms\ channel maps is 66~mJy/beam.

\subsection{IRAM 30-m telescope}
\label{ssd}

Maps of G31.41 with the 30-m telescope were made on November 4th, 2007. The
HERA multi-beam receiver was used to cover a region of
$2\arcmin\times2\arcmin$ centered on
$\alpha$(J2000)=18$^{\rm h}$47$^{\rm m}$34\fs3 and
$\delta$(J2000)=--01\degr12\arcmin45\farcs9,
using the on-the-fly-mode. To prevent systematic effects, the region was
scanned alternatively along right ascension and declination, and all the data
were eventually averaged to obtain the final map. The receiver was tuned to
the frequencies of the \CO(2--1) (HERA1) and \COI(2--1) (HERA2) lines, which
were hence observed simultaneously. Both lines were covered with the VESPA
autocorrelator, with 0.4~\kms\ spectral resolution.
The reference position used for all maps was 
$\alpha$(J2000)=18$^{\rm h}$45$^{\rm m}$15\fs24 and
$\delta$(J2000)=--00\degr55\arcmin56\farcs09. This was carefully chosen
from CO surveys such as that by Sanders et al. (\cite{sand}), and checked to
be free of emission in the observed lines.
Data were reduced with the program CLASS of the GILDAS package\footnote{
The GILDAS software is available at http://www.iram.fr/IRAMFR/GILDAS
}
and channel maps with 1~\kms\ resolution were created.

\section{Results}
\label{sres}

In the following we illustrate the results obtained, first for the typical
hot core tracers (continuum, \MCN, and \MCNII\ lines\footnote{
Parameters of the \MCN\ and \MCNII\ transitions can be found, e.g., in the
``splatalogue'' database: http://www.splatalogue.net/
}
) and then for the
outflow tracers (\CO\ and isotopomers).

\subsection{Core tracers}
\label{score}

Compared to previous Plateau de Bure interferometer (PdBI) observations
in the same lines (BEL04, BEL05), our SMA
observations benefit from a better uv coverage and more circular beam, thus
allowing us to attain sub-arcsec resolution in all directions (see
Table~\ref{tobs}; for comparison, the synthesized beam of the PdBI images was
1\farcs1$\times$0\farcs5 with PA=189\degr).

\begin{figure}
\centering
\resizebox{8.5cm}{!}{\includegraphics[angle=0]{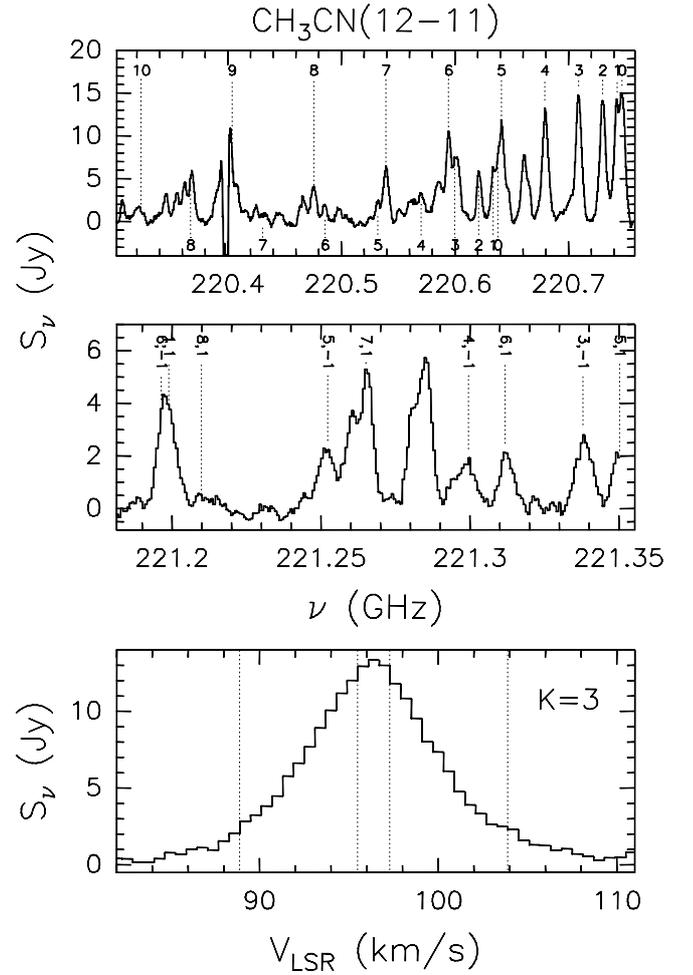}}
\caption{
Spectra obtained by integrating the line emission over the whole core.
{\bf Top.} Ground-state $K$-components of the $J$=12$\rightarrow$11 transition
 of \MCN\ (numbers above the spectrum) and of the \MCNII\ isotopomer (numbers
 below the spectrum). Note how most $K$-lines are blended with each other and/or
 with lines of other molecules.
{\bf Middle.} Vibrationally excited ($v_8$=1) lines of \MCN. The $K$,$l$
 quantum numbers of the transitions are indicated above the spectrum.
{\bf Bottom.}
Ground state $K$=3 component. The dotted lines mark the velocity ranges
used to produce the maps in Figs.~\ref{fmapsu} and~\ref{fmapsd}.
}
\label{fmcnsp}
\end{figure}

To give an idea of the complexity of the line emission from the G31.41
HMC, in Fig.~\ref{fmcnsp} we show a spectrum covering all \MCN(12--11)
transitions with $K\le10$, obtained by integrating the emission over the
whole core.
For an estimate of the \MCN\ and \MCNII\ line parameters we refer to the study
of BEL05 (see their Table~4).
We did not attempt the identification of all lines detected,
because this goes beyond the purposes of the present study. However, we note that
a number of \MCN\ and \MCNII\ transitions are heavily blended with each other
and/or with other lines. Only the \MCN\ $K$=2,3,4,8 and the \MCNII\ $K$=2,6
components appear free enough of contamination to be used for a detailed
investigation. A similar consideration
holds for the vibrationally excited ($v_8$=1) lines, shown in the same
figure. Here, only the ($K$=3,\,$l$=--1) and ($K$=6,\,$l$=1) transitions are
clearly detected and sufficiently separated from other lines to be
considered for our study. Note that our bandwidth covers all $v_8$=1
lines down to the (11,--1) at 220788.016~MHz, but the lines below 221.19~GHz
are too weak or too blended: this is the reason why the \MCN(12--11) $v_8$=1
spectrum in Fig.~\ref{fmcnsp} is displayed only above this frequency.

\begin{figure*}
\centering
\resizebox{14cm}{!}{\includegraphics[angle=0]{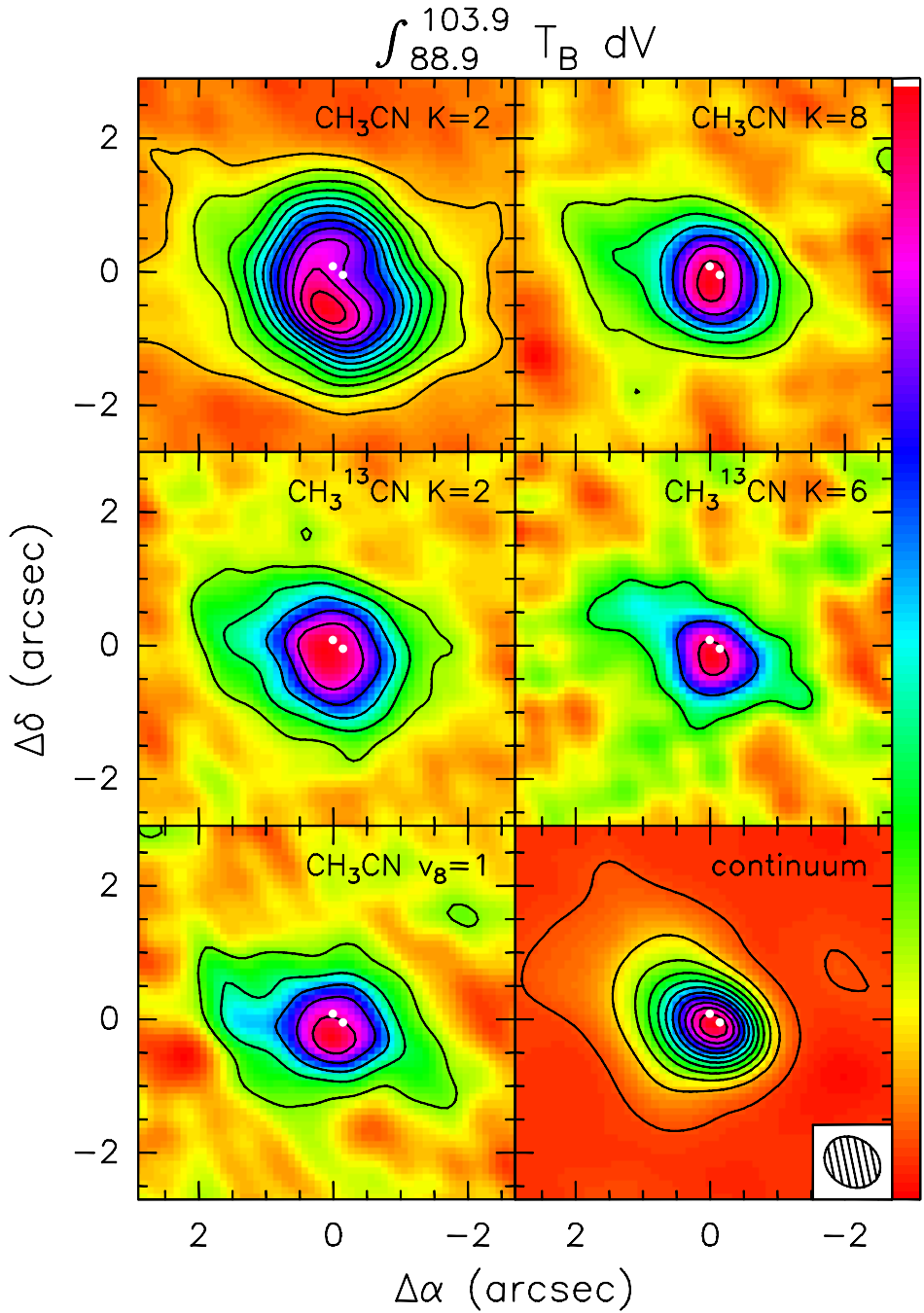}}
\caption{
Maps of the line and continuum emission toward the G31.41 HMC. The line
maps have been obtained by averaging the emission under the whole line
profile, namely from 88.9~\kms\ to 103.9~\kms\ 
(see Fig.~\ref{fmcnsp}, bottom).
The tracer used is indicated
in the top right of each panel. The two white points mark the positions of
the free-free-continuum sources detected by Cesaroni et al.~(\cite{cesa10}).
The contour levels start from 0.072~Jy/beam and increase in steps of
0.12~Jy/beam for all line maps, and range from 0.035~Jy/beam to 1.785~Jy/beam
in steps of 0.175~Jy/beam for the continuum map.
Offsets are measured with respect to the phase center
$\alpha$(J2000)=18$^{\rm h}$47$^{\rm m}$34\fs315
$\delta$(J2000)=--01\degr12\arcmin45\farcs9. The ellipse in the bottom right
denotes the FWHP of the synthesized beam.
}
\label{fmapsu}
\end{figure*}

\begin{table}
\begin{flushleft}
\caption[]{Continuum parameters, i.e. peak position, peak brightness
temperature in the synthesized beam, integrated flux density,
full width at half maximum (FWHM), and deconvolved angular diameter
}
\label{tcont}
\begin{tabular}{cccccc}
\hline
$\alpha$ & $\delta$ & $T_{\rm B}$ & $S_\nu$ & FWHM & $\Theta$ \\
(J2000)  & (J2000)  &  (K)        & (Jy)    & (arcsec) & (arcsec) \\
\hline
18$^{\rm h}$47$^{\rm m}$34\fs31 & --01\degr12\arcmin46\farcs0 & 68 & 4.6 & 1\farcs1 & 0\farcs76 \\
\hline
\end{tabular}
\end{flushleft}
\end{table}

\begin{figure*}
\centering
\resizebox{14cm}{!}{\includegraphics[angle=0]{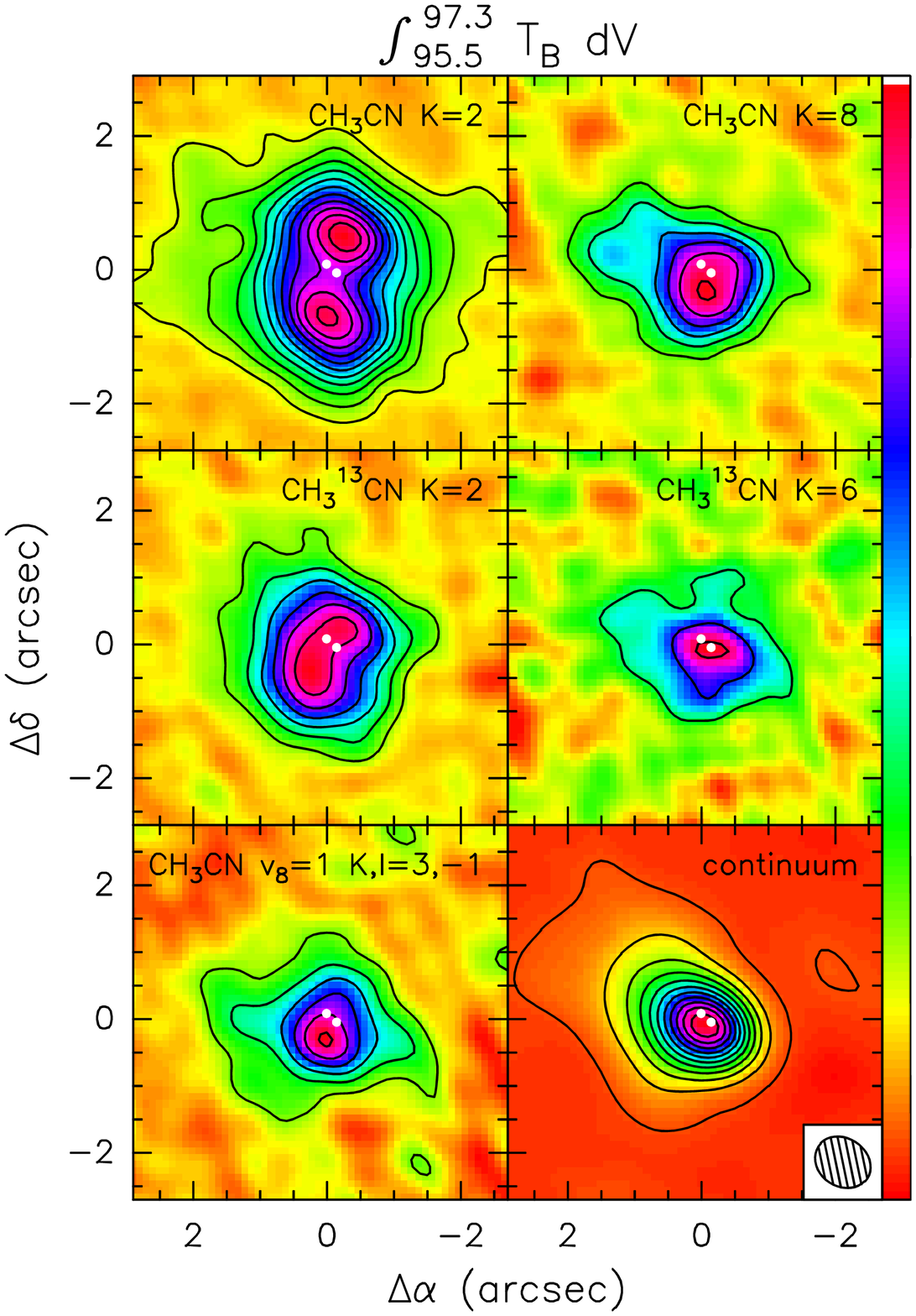}}
\caption{
Same as Fig.~\ref{fmapsu}, with the line maps obtained by averaging the
emission from 95.5~\kms\ to 97.3~\kms\ 
(see Fig.~\ref{fmcnsp}, bottom).
Contour levels start from
0.15~Jy/beam and increase in steps of 0.2~Jy/beam for all line maps, and
range from 0.035~Jy/beam to 1.785~Jy/beam in steps of 0.175~Jy/beam for the
continuum map.
}
\label{fmapsd}
\end{figure*}

Figure~\ref{fmapsu} shows continuum and line emission maps obtained by
averaging the emission under the line and Table~\ref{tcont} lists the main
parameters of the continuum emission. The most obvious feature
is that the maps of all tracers are slightly elongated
in the NE--SW direction, suggesting that the distribution of the molecular
gas is flattened approximately in the direction defined by the two
free-free continuum sources detected by Cesaroni et al.~(\cite{cesa10}).

From the same figure one can also note that the high-energy transitions and
the isotopomer emission peak approximately at the same position as the
continuum, whereas the maximum of the low-energy $K$=2 component appears to be
slightly offset from it. Such an offset is evident for all components with
$K\le6$, while the $K$=7 and~8 line maps look more circular and peak at
the core center. Nothing can be said about the $K$=9 transition,
which is heavily blended with the \COI(2--1) line. A plausible interpretation
is that the opacity decreases with increasing excitation energy. In
confirmation of that, one can see that this effect is even more prominent in
the maps obtained by integrating the line emission over a narrow velocity
range around the systemic LSR velocity, i.e. from 95.5 to 97.3~\kms (see
Fig.~\ref{fmapsd}). Now, the $K$=2 emission clearly splits into two peaks
located on opposite sides with respect to the continuum and also the
corresponding \MCNII\ line map shows a similar pattern.
We conclude that opacity plays a crucial role in the \MCN\ emission of
this HMC and cannot be neglected in the analysis of the line emission.

\begin{figure*}
\centering
\resizebox{13.5cm}{!}{\includegraphics[angle=-90]{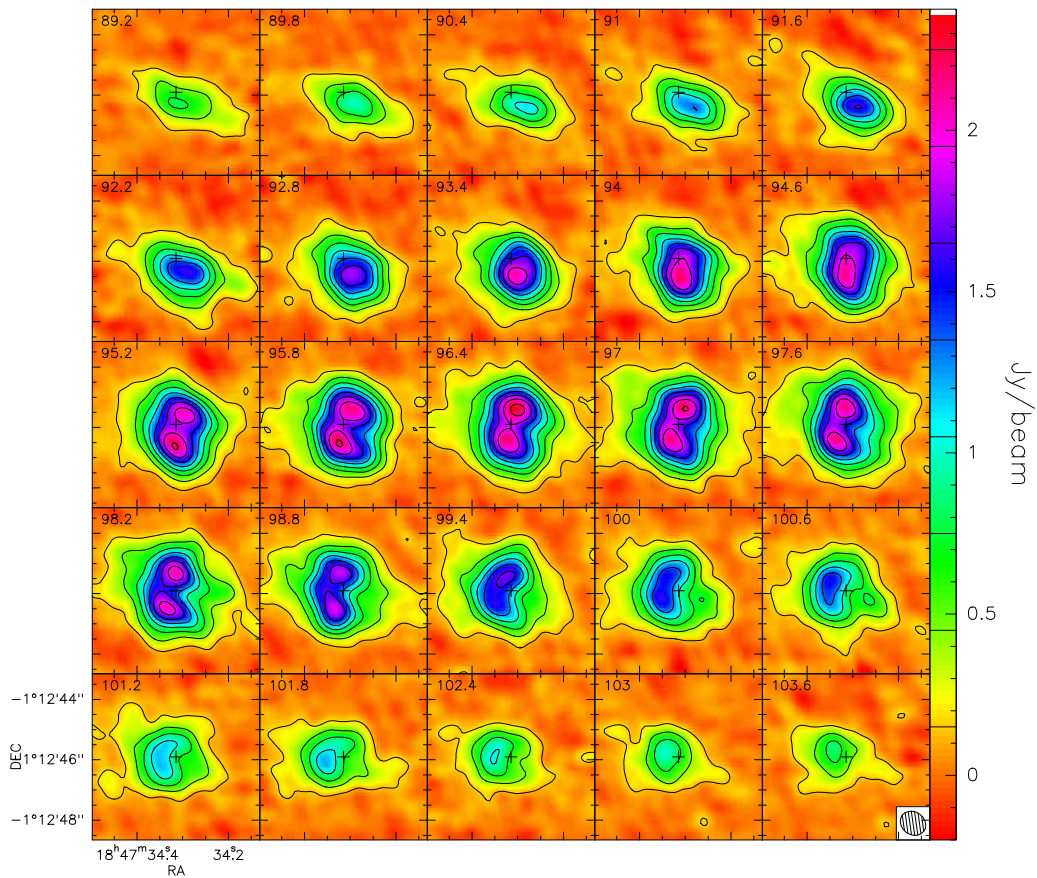}}
\caption{
Channel maps of the \MCN(12--11) emission obtained by averaging the $K$=2, 3,
and 4 components.
The number in the top left of each panel
gives the corresponding LSR velocity. The values of the contour levels are
indicated by the tick marks in the intensity scale to the right.
The cross denotes the phase center.
The ellipse in the bottom right of the
last panel is the full-width-at-half power of the synthesized beam.
}
\label{fmcnchm}
\end{figure*}

A qualitative impression of the \MCN\ emission can be obtained from visual
inspection of the channel maps shown in Fig.~\ref{fmcnchm}. In order to
improve the signal-to-noise, such maps have been obtained by averaging the
emission in the $K$=2, 3, and 4 components. These transitions have been chosen
because they do not seem to be affected by significant contamination by other
lines and have comparable excitation energies, which suggests that they
are likely tracing the same gas.

\begin{figure}
\centering
\resizebox{8.5cm}{!}{\includegraphics[angle=-90]{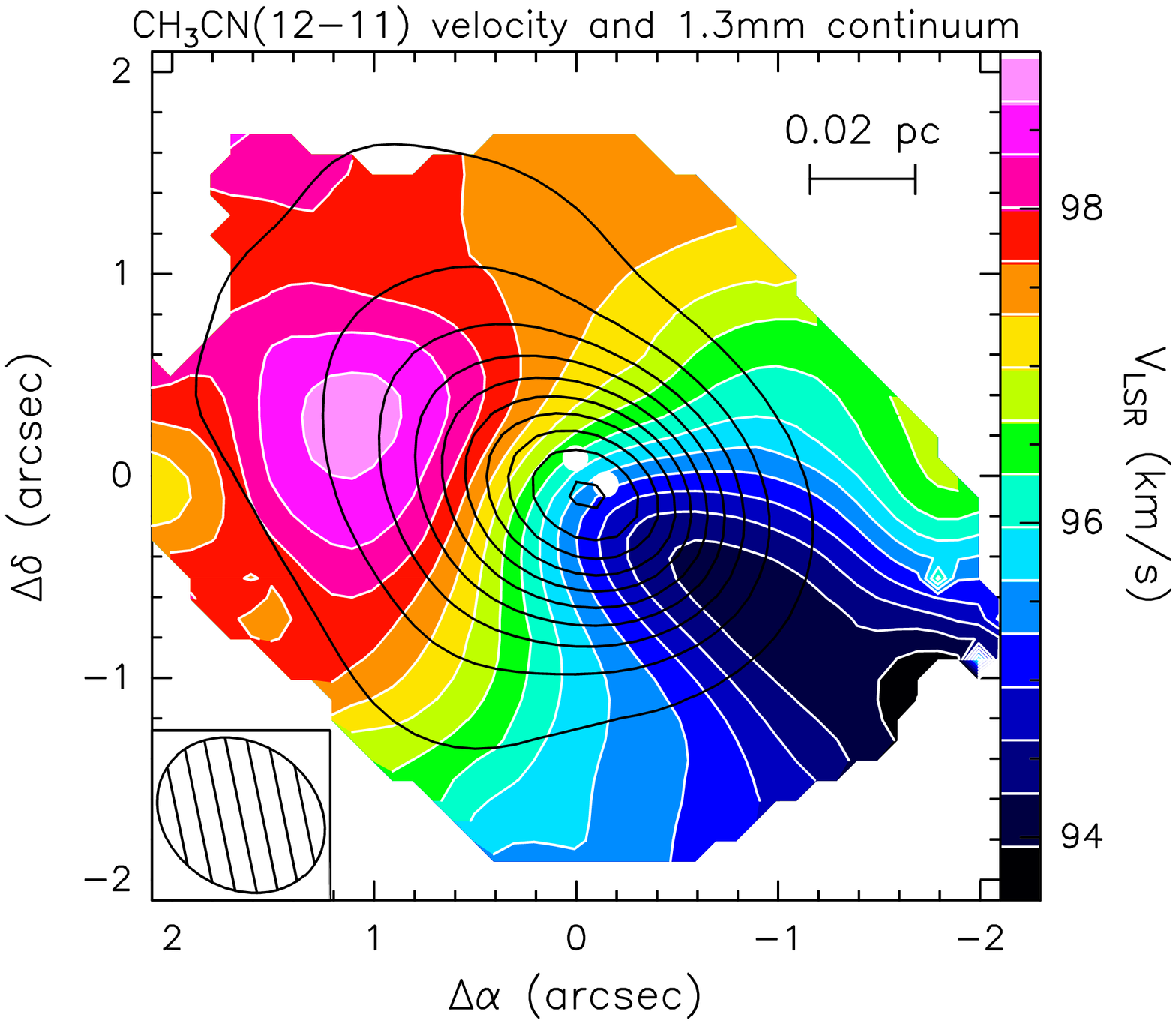}}
\caption{
Overlay of the map of the 1.3~mm continuum emission (contours) on that
of the \MCN(12--11) line velocity (color scale). Contour levels range from
0.1 to 2 in steps of 0.2~Jy/beam. Offsets are measured with respect to the
phase center.
The two white dots denote the free-free continuum sources detected
by Cesaroni et al.~(\cite{cesa10}).
The synthesized beam is shown in the bottom left.
}
\label{fvmap}
\end{figure}

\begin{figure*}
\centering
\resizebox{14cm}{!}{\includegraphics[angle=0]{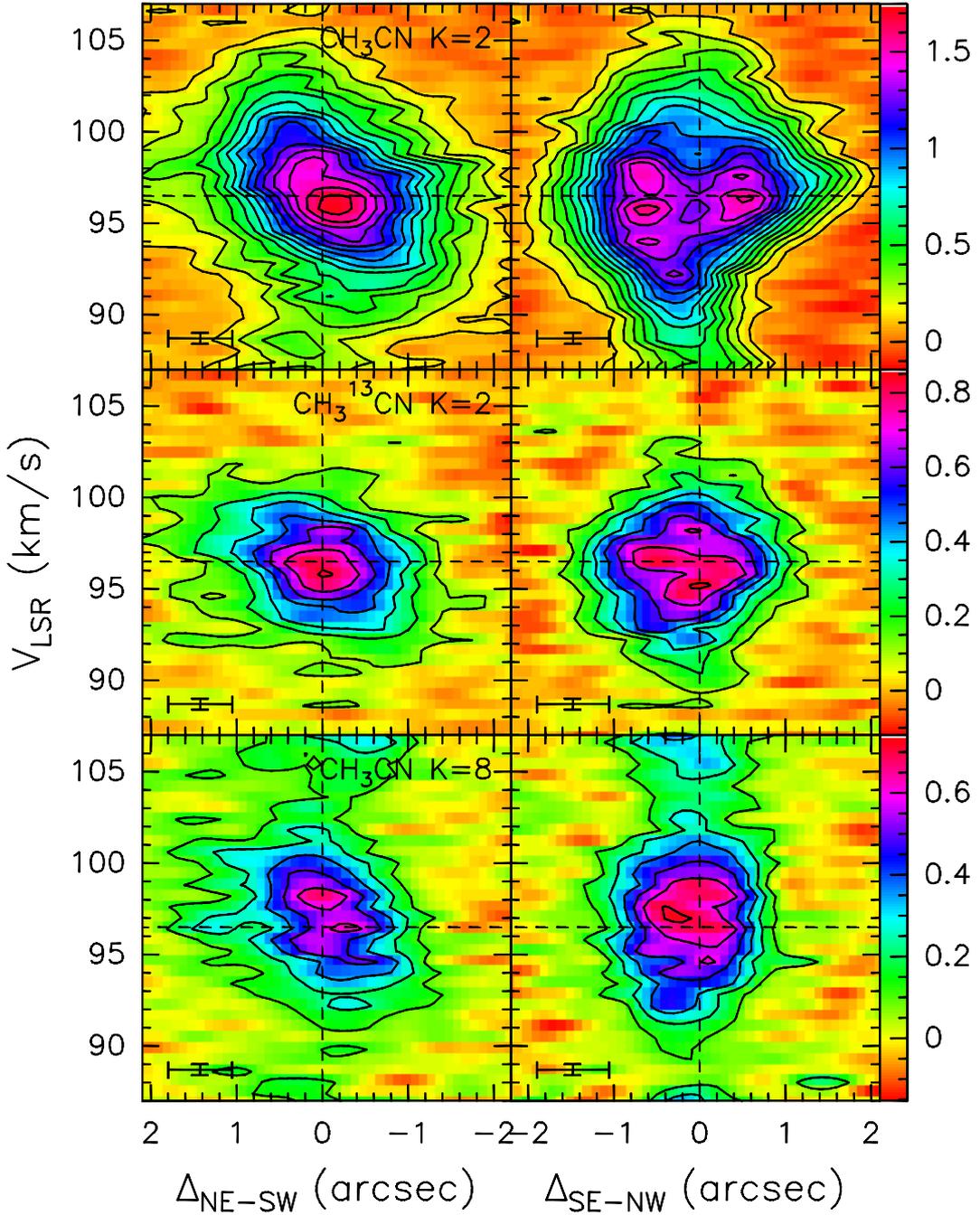}}
\caption{
Position--velocity plots of along P.A.=68\degr\ (left panels) and --22\degr\
(right panels) for three different lines (indicated in each panel).
Offsets are measured with respect to the
phase center. Contour levels start from 0.12~Jy/beam and increase in steps
of 0.12~Jy/beam. 
The cross in the bottom left of each box denotes the angular and spectral
resolutions.
}
\label{fpvplots}
\end{figure*}

Clearly, the emission gradually drifts from SW to NE with increasing velocity
and outlines an ``8-shaped'' pattern close to the systemic velocity of
$\sim$96.5~\kms. These characteristics resemble those observed in rotating
circumstellar disks around low-mass stars (see e.g. Fig.~1 of Simon et
al.~\cite{simon}). We will discuss this problem in Sect.~\ref{sotto}.
A more quantitative analysis of the velocity field traced by the \MCN\ emission
is attained -- in analogy with the previous study by BEL04 --
by fitting the $K$=0 to~4
components simultaneously, after fixing their separations in frequency to the
laboratory values and forcing the line widths to be equal\footnote{
For this purpose, program CLASS of the GILDAS package was used, after
extracting the relevant spectra from the data cube and converting them
to the appropriate format.
}.
This procedure
improves the quality of the fit
with respect to fitting a single $K$ line,
but we stress that a similar result is
obtained by fitting a single Gaussian to any of the unblended $K$ components.
A map of the LSR
velocity is presented in Fig.~\ref{fvmap}, which confirms the existence of a
clear velocity gradient directed approximately NE--SW (P.A.$\simeq$68\degr)
and roughly centered on the peak of the continuum emission. A different
and more detailed representation of the same velocity gradient is
presented in Fig.~\ref{fpvplots}, where the position--velocity plots along
the velocity gradient (P.A.=68\degr) and perpendicular to it (P.A.=--22\degr)
are shown for three different lines of \MCN\ and \MCNII.
Note that to enhance the signal-to-noise ratio in the plots, we
have averaged the emission along the direction perpendicular to the cut
along which the position--velocity plot is calculated.
As expected,
no velocity trend is seen in the plots along P.A.=--22\degr.

\begin{figure}
\centering
\resizebox{8.0cm}{!}{\includegraphics[angle=-90]{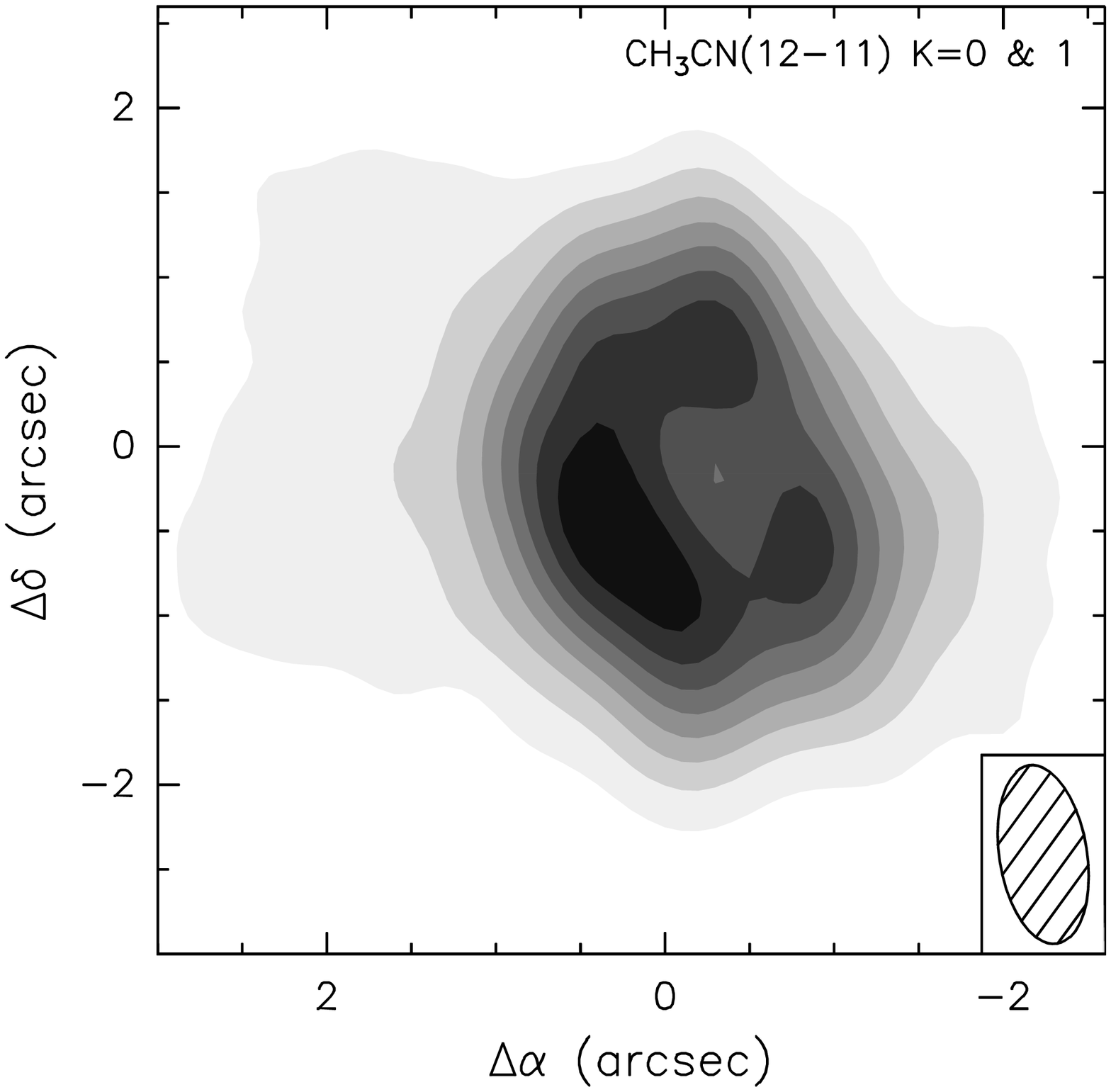}}
\caption{
Map of the \MCN(12--11) emission obtained by averaging the emission under the
$K$=0 and~1 components. The image has been reconstructed with a clean beam
(ellipse in the bottom right)
equal to that of the images of BEL04, i.e.
1\farcs1$\times$0\farcs5 with PA=189\degr. The contour levels range from
75 to 600 in steps of 75~mJy/beam. Offsets are measured with respect to the
phase center.
}
\label{fpdb}
\end{figure}

It is worth comparing our results to those of BEL04.
With respect to their Fig.~2e, our Fig.~\ref{fvmap} presents
a slightly broader velocity range ($\sim$5.5~\kms\ instead of $\sim$4~\kms),
over a larger region ($\sim$3\arcsec\ instead of $\sim$2\arcsec). These
differences are probably caused by the different angular resolutions and
sensitivities.
A more evident discrepancy appears in the map of the \MCN(12--11) emission,
i.e. our Fig.~\ref{fmapsd} and their Fig.~1. In the
former, the $K$=2 emission appears to outline two peaks located approximately to
the NW and SE with respect to the core center, whereas in Fig.~1 of BEL04
the peaks lie to the E and W. However, one should
keep in mind that this map was obtained integrating the emission under the
$K$=0,1, and 2 components. Moreover, we have averaged the emission over a
narrow interval around the peak, whereas BEL04 integrated under the whole
line profile. Finally,
the angular resolution is different for the two data sets. In order to
attain a more consistent comparison, we have reconstructed our maps
using the same clean beam as BEL04 and integrating over the same
frequency interval. The new map (shown in Fig.~\ref{fpdb})
is now much more similar to that of BEL04, demonstrating how
the apparent morphology of the emission may depend significantly on the
resolution. Given the larger number of antennas and more
circular beam, we believe that our SMA images reproduce the
structure of the emission more faithfully than the old maps by BEL04 and BEL05.

\subsection{Outflow tracers}
\label{soutf}

\begin{figure*}
\centering
\resizebox{14cm}{!}{\includegraphics[angle=-90]{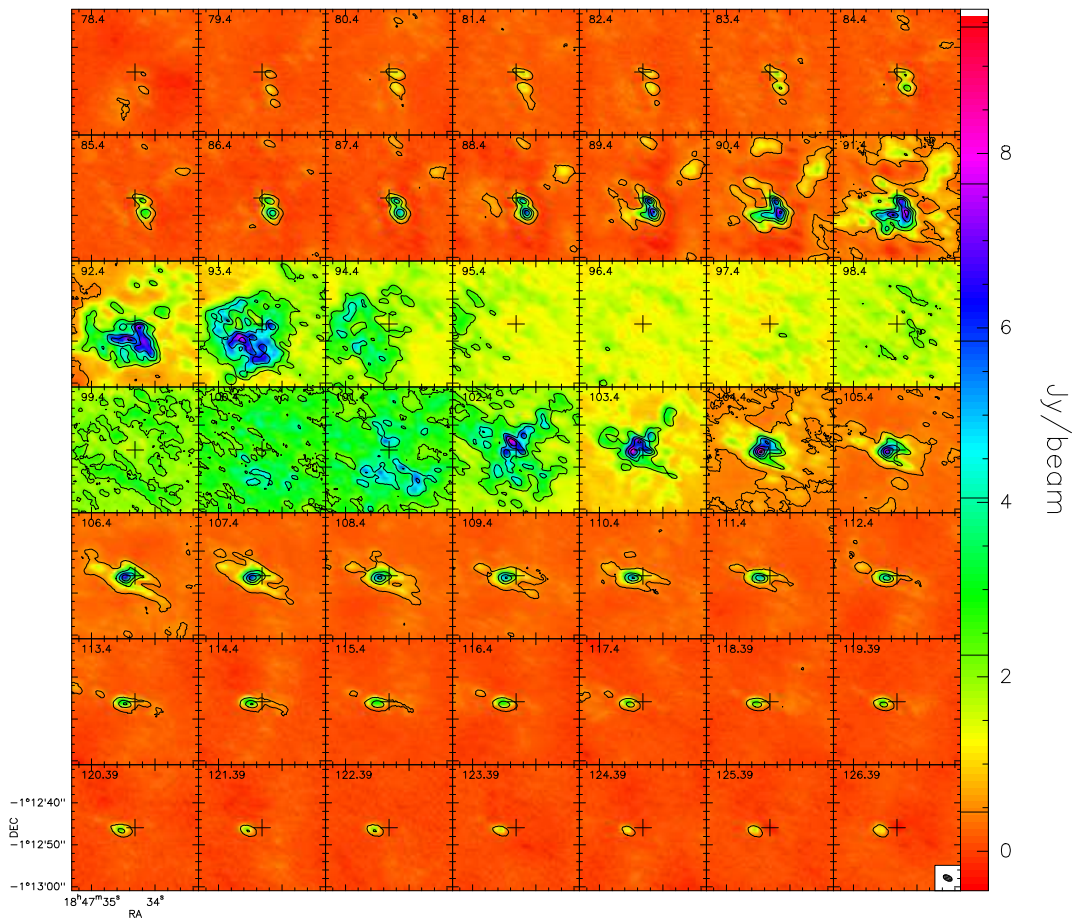}}
\caption{
Channel maps in the \CO(2--1) line obtained by combining single-dish (IRAM 30-m)
and interferometric (SMA) data. The number in the top left of each panel
gives the corresponding LSR velocity. The values of the contour levels are
indicated by the tick marks in the intensity scale to the right.
The ellipse in the bottom right of the
last panel is the full-width-at-half power of the resulting instrumental beam.
}
\label{fchm12co}
\end{figure*}

One of the purposes of the present study was to compare the structure and
kinematics of the HMC with those of a possible bipolar outflow associated
with it. The existence of such an outflow had been suggested by
Olmi et al.~(\cite{olmi96b}), whose Plateau de Bure images in the \COI(1--0)
line seem to reveal two narrow lobes oriented SE--NW (see their Fig.~5).
With this in mind, we have imaged G31.41 in the \CO\ and \COI\ $J$=2--1
rotational transitions with both the SMA and 30-m IRAM telescope, to recover
also the emission filtered out by the interferometer.
Figure~\ref{fchm12co}
shows channel maps in the \CO(2--1) line and effectively
demonstrates the complexity of the region. Only at high velocities the
emission appears quite compact, whereas close to the systemic velocity
the structure presents a complicated pattern. In particular, from 95 to
100~\kms\ the emission is uniformly distributed over an extended region:
this means that the line opacity is sufficiently high that one can see
only the surface of the cloud at these velocities, consistent with the
relatively low values of the brightness temperature (10--15~K).

One arrives at the same conclusion by looking at Fig.~\ref{fcospts}, where
a comparison is shown of the single-dish spectra obtained by averaging
the emission of three different CO isotopomers over a square region,
30\arcsec\ in size, centered at the HMC position. While the \CO\ and
\COI\ profiles are double-peaked with a dip at $\sim$97~\kms, the optically
thin(ner) \COIII(2--1) line (Cesaroni, unpublished data) is Gaussian and
peaks right at the velocity of the dip in the other two lines. This is an
indication of self-absorption and thus of high optical depth in the \CO\ and
\COI\ transitions.

An apparent feature of the \CO(2--1) profile is the presence of broad wings.
From Fig.~\ref{fchm12co} one sees that this high-velocity emission originates
from 2--3 compact structures located roughly to the E (red-shifted) and
W (blue-shifted) of the HMC. Whether this morphology is consistent with
that of a bipolar outflow cannot be trivially decided on the basis of the
evidence presented so far and requires a detailed analysis and discussion
that we postpone to Sect.~\ref{svgr}.

\section{Discussion}
\label{sdis}

The main purpose of the present study is to shed light on the nature of the
velocity gradient observed in this HMC (e.g. Fig.~\ref{fvmap}). As already
mentioned, BEL04 and BEL05 favored the
rotating toroid scenario. In contrast, other authors (Gibb et al.~\cite{gibb};
Araya et al.~\cite{araya}) preferred the outflow interpretation, thus posing a
problem that we wish to address with our new SMA data.

Besides the (possible) existence of rotation and/or outflow, the situation in
G31.41 is complicated by the existence of infall, detected by Girart et al.
(\cite{gira}) as an inverse P-Cygni profile of the \CSI(7--6) line observed
with the SMA. Therefore, before making a comparative discussion of the
different models for the velocity gradient, we need to verify to what extent
our \MCN\ measurements may be affected by infall.

\subsection{Evidence for infall}

\begin{figure}
\centering
\resizebox{8.5cm}{!}{\includegraphics[angle=-90]{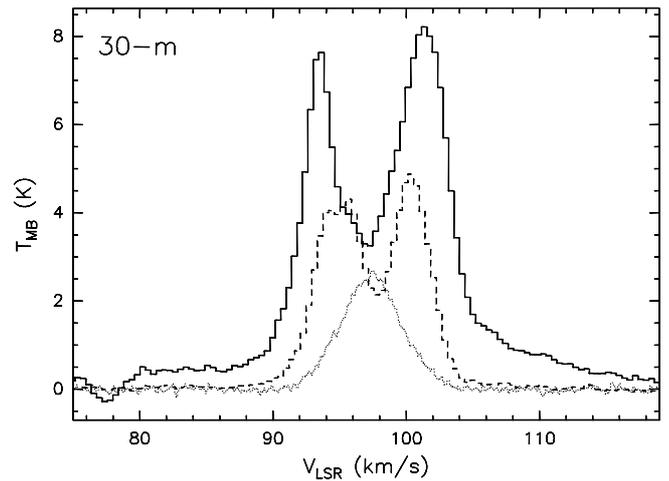}}
\caption{
Spectra of the \CO(\ (solid histogram), \COI\ (dashed), and \COIII\ (dotted;
Cesaroni, unpublished data) (2--1) lines obtained by averaging the emission
over a square region 30\arcsec\ in size, centered on the HMC. The data have
been taken with the IRAM 30-m telescope.
}
\label{fcospts}
\end{figure}

As already discussed in Sect.~\ref{soutf}, Fig.~\ref{fcospts} clearly reveals
that the \CO\ and \COI\ (2--1) lines are affected by self-absorption. However,
only a weak asymmetry is seen in the line profile, with the red-shifted peak
being slightly more prominent than the blue-shifted one. If such an asymmetry
were caused by infalling gas, the blue-shifted peak should be stronger;
hence we can conclude that no obvious evidence of infall is present on the
large scale (30\arcsec\ or 1.1~pc) over which the spectra in
Fig.~\ref{fcospts} have been calculated. It thus seems that the infall
detected by Girart et al.~(\cite{gira}) occurs close to the
HMC. Do we see any evidence of this in the HMC tracers that we observed?
All \MCN\ and \MCNII\ (12--11) lines shown in Figure~\ref{fmcnsp} have
Gaussian profiles with no hint of (self-)absorption, unlike the \CSI(7--6)
line observed by Girart and collaborators. However, this is not surprising,
because both the excitation energy and critical density of the latter (49~K
and $\sim10^7$~\cmc) are less than those of the former transitions (58--926~K
and $\ga10^8$~\cmc). This implies that the \CSI(7--6) line traces large
radii of the HMC that are characterized by a relatively low temperature, and thus
absorb the hot continuum photons emitted from the central region of the
core. This effect does not apply to the \MCN(12--11) transitions, because
these arise from a smaller hot shell, whose excitation temperature is much
more similar to the brightness temperature of the dust continuum.

\begin{figure}
\centering
\resizebox{8.5cm}{!}{\includegraphics[angle=0]{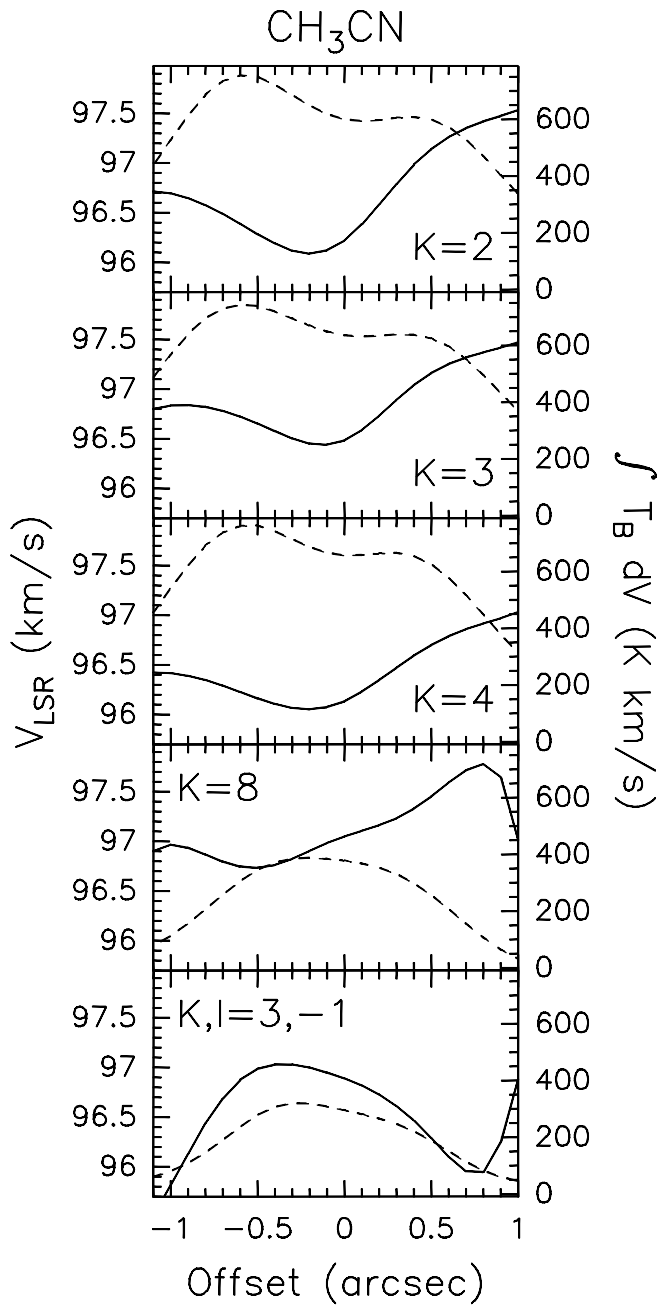}}
\caption{
Solid curves represent the first moments of several \MCN(12--11)
components (indicated in the top left of each panel)
estimated along a cut through the continuum peak
($\alpha$(J2000)=18$^{\rm h}$47$^{\rm m}$34\fs31 $\delta$(J2000)=--01\degr12\arcmin46\farcs0) 
with P.A.=--22\degr, i.e. perpendicular to the velocity gradient in
Fig.~\ref{fvmap}. The dashed curves are the zero moments of the corresponding
lines.
}
\label{fperp}
\end{figure}

Despite the apparent lack of evidence for infall in the \MCN\ transitions,
we have attempted a more detailed analysis to see whether some hint of
asymmetry was present in the line profiles. The idea is that if the line
shape is skewed toward the blue (although very weakly), a Gaussian fit
should give a peak velocity less than the systemic velocity. With this in mind,
in Fig.~\ref{fperp} we plot the line first moment (solid curves) as a
function of distance from the HMC center along an axis with P.A.=--22\degr,
i.e. perpendicular to the direction of the velocity gradient. For the sake of
completeness, also the line zero moment (integrated intensity) along the same
direction is shown (dashed curves). This is done for the $K$=2,3,4, and 8
components of the ground-state transition and the (3,-1) line of
\MCN\ $v_8$=1. Evidently, up to $K$=8 the line velocity attains a
minimum value close to the HMC center, whereas this minimum is not seen in
the $v_8$=1 line. This proves that the lower energy lines are slightly skewed
to the blue toward the bright continuum peak, as expected for weak
red-shifted absorption caused by infall. This effect is not seen in the highest
energy line, coming from the innermost, hottest layers of the core.

We conclude that some of the \MCN(12--11) lines are
affected by red-shifted absorption, although in a much less
prominent way than the \CSI(7--6) line measured by Girart et al.
(\cite{gira}). Note that this absorption causes only a marginal shift in the
line velocity, of $\sim$0.5~\kms, and is going to have only negligible
effects on our study of the velocity shift of $\sim$5~\kms\ observed
across the HMC.

\subsection{The NE--SW velocity gradient}
\label{svgr}

The existence of a velocity gradient in the \MCN\ lines is proved beyond any
doubt by Figs.~\ref{fmcnchm}, \ref{fvmap}, and~\ref{fpvplots}. The high
angular resolution allows us to establish that the velocity is
shifting smoothly across the core, and one can reasonably exclude that this
shift is caused by two unresolved velocity components (sub-cores) with different
velocities. The question we intend to address here is whether this velocity
gradient
is caused by rotation of the HMC about a SE--NW axis, as hypothesized by BEL04,
or to expansion in a bipolar outflow oriented NE--SW, as proposed by Gibb et
al.~(\cite{gibb}) and Araya et al.~(\cite{araya}). How can one distinguish
between these two scenarios? As explained in Sect.~\ref{sint},
it is common belief that disks and outflows are tightly associated,
with the latter being ejected along the rotation axis of the former.
Therefore, if the velocity gradient in the G31.41 HMC is caused by rotation,
one would expect to detect a bipolar outflow perpendicular to it, on
a larger scale. In contrast, if the velocity gradient is tracing the ``root''
of a bipolar outflow, on a larger scale this outflow should
become clearly visible along the same direction defined by the (small-scale)
velocity gradient. Our combined single dish and
interferometric observations in typical outflow tracers such as \CO\ and
\COI\ should hence be well suited for our purposes, given the sensitivity to
both small (a few arcsec) and large (a couple of arcmin) scales.

\begin{figure*}
\centering
\resizebox{14cm}{!}{\includegraphics[angle=0]{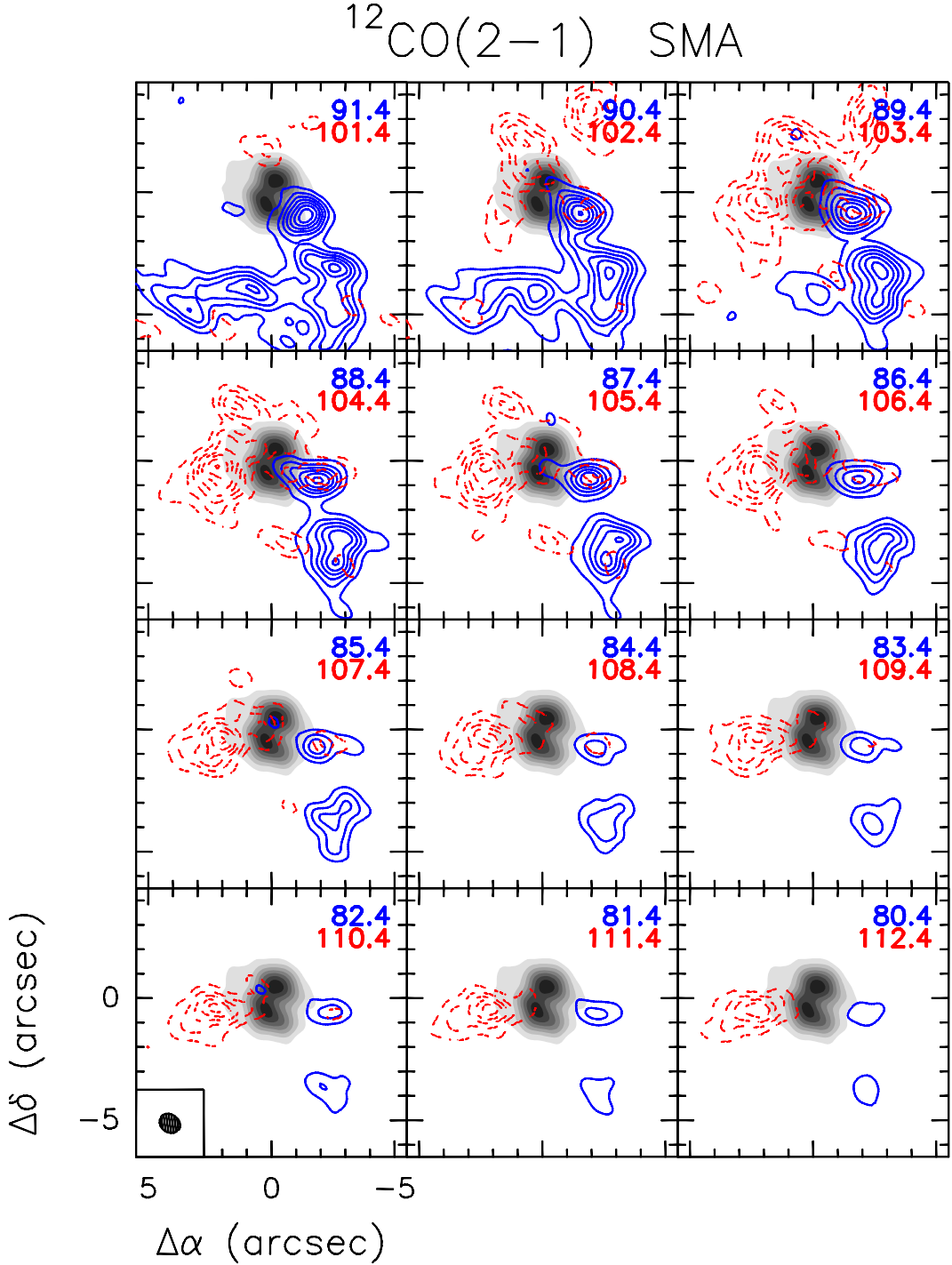}}
\caption{
Channel maps of the \CO(2--1) line emission observed with the SMA.  Each box
contains pairs of maps corresponding to the blue- (solid contours) and
red-shifted (dashed) emission at the same velocity offset (in absolute value)
from the the systemic velocity ($\sim$96.5~\kms). The corresponding LSR
velocities are indicated in the top right of the box. Contour levels increase
from 0.5~Jy/beam in steps of 0.5~Jy/beam. Also shown is a map of the
\MCN(12--11) $K$=4 line emission averaged over the range 95.2--96.7~\kms.
The ellipse in the bottom left denotes the synthesized beam of the
observations.
}
\label{fcorb}
\end{figure*}

The results reported in Sect.~\ref{soutf} are quite ambiguous.
The blue- and red-shifted \CO\ and \COI\ gas is oriented like the \MCN\
velocity gradient, suggesting that the two are manifestations of the same
phenomenon. The existence of a NE--SW bipolarity in the \CO\ emission
can be appreciated in Fig.~\ref{fcorb}, where we show the blue- and
red-shifted \CO(2--1) emission in pairs of channels equally offset from the
systemic velocity. One is tempted to conclude that \CO\ is indeed associated
with a bipolar outflow, whose densest component is traced by the \MCN\ emission.
However, this bipolarity in the CO maps is seen only on a small scale, because no
evidence of high-velocity CO emission is found beyond $\sim$5\arcsec\ (i.e.
0.2~pc) from the HMC, implying an unusually small size for a typical outflow
associated with a high-mass star-forming region. This casts some doubt on the
outflow interpretation. Below we discuss the two hypotheses (outflow
and toroid) in detail.

\subsubsection{Bipolar outflow}

The outflow hypothesis is supported by a couple of facts. First of all,
if the \MCN\ gradient is caused by a rotating toroid, one expects to find
a bipolar outflow perpendicular to it, but this is not seen in our maps.
Second, the position velocity plot of the \CO\ emission in the direction
of the \MCN\ velocity gradient suggests that the gas velocity is proportional
to the distance from the star (as discussed later in Sect.~\ref{storo}),
consistent with the Hubble-law expansion observed in molecular outflows from
YSOs.

In order to check the plausibility of the outflow scenario, we have
calculated the outflow parameters from the \MCN,
\MCNII, \CO, and \COI\ lines, in the latter two cases using the combined 30-m
and SMA data.
These are given in Table~\ref{tout}, where we report the mass of the outflow,
$M$, the momentum, $P$, the energy, $E$, and the corresponding rates obtained
by dividing the previous quantities by the dynamical time scale (see below).
In practice, $M=\sum_i m_i$, $P=\sum_i m_i V_i$, and $E=\sum_i (1/2) m_i
V_i^2$, where the sums are extended over only those channels, $i$, falling in
the velocity ranges given in the footnotes of Table~\ref{tout}, and $m_i$ is
the mass moving with velocity $V_i$ relative to the systemic velocity. This
mass is computed by integrating the line emission inside the regions
corresponding to the $5\sigma$ level of the blue- and red-shifted emission.
Note that
the velocity intervals were chosen by inspecting the channel maps
and selecting only those channels where the emission was sufficiently strong
and at the same time not affected by the missing flux problem close to the
systemic velocity.
In our calculations we assume a temperature of $\sim$100~K for
both molecules, intermediate between the peak brightness temperature (77~K)
of the \CO(2--1) line and the rotational temperature (164~K) estimated by
BEL05 from the \MCNII(6--5), \MCNII(12--11), and \MCN(6--5)~$v_8$=1 lines
(see their Fig.~6). While this temperature may seem too high for the outflow
component traced by the CO isotopomers, assuming 50~K instead of 100~K
would reduce the outflow parameters by only a factor 0.6.
The abundance of \MCN\ and \CO\ relative to \HM\ are
assumed equal to $10^{-8}$ and $10^{-4}$ respectively (see e.g. Van Dishoek
et al. \cite{vand}), while the isotopic ratios \CO/\COI\ and \MCN/\MCNII\ are
taken equal to 50 after Wilson \& Rood (\cite{wiro}), for a galactocentric
distance of 4.5~kpc. The dynamical time scale of the outflow,
$t_{\rm dyn}\simeq4\times10^3$~yr, is calculated from the maximum size
($\sim$0.12~pc) and velocity ($\sim$30~\kms) of the \CO\ lobes. Note that the
(unknown) inclination of the outflow with respect to the line of sight has
not been taken into account in our estimates.

\begin{table}
\begin{flushleft}
\caption[]{Outflow parameters calculated from different lines. The dynamical
time scale is $t_{\rm dyn}\simeq4\times10^3$~yr and a gas temperature of 100~K
is assumed in the calculations}
\label{tout}
\begin{tabular}{ccccc}
\hline
parameter & \CO\ $^a$ & \COI\ $^b$ & \MCN\      & \MCNII\    \\
          &           &            & $K$=4 $^c$ & $K$=2 $^d$ \\
\hline
$M (M_\odot)$ & 3.8 & 20 & 60 & 290 \\
$P (M_\odot\,\kms)$ & 48 & 230 & 350 & 1200 \\
$E (L_\odot\,{\rm yr})$ & $5.9\times10^4$ & $2.2\times10^5$ & $1.7\times10^5$ & $4.2\times10^5$ \\
$\dot{M} (M_\odot\,{\rm yr}^{-1})$ & $9.5\times10^{-4}$ & $5\times10^{-3}$ & 0.015 & 0.07 \\
$\dot{P} (M_\odot\,\kms{\rm yr}^{-1})$ & 0.012 & 0.057 & 0.087 & 0.30 \\
$\dot{E} (L_\odot)$ & 14.7 & 54.6 & 41.8 & 104 \\
\hline
\end{tabular}

\vspace*{1mm}
$^a$~blue wing from 78.9 to 89.9~\kms, red wing from 105.9 to 127.9~\kms. Abundance relative to \HM: $10^{-4}$ \\
$^b$~blue wing from 77.9 to 88.9~\kms, red wing from 103.9 to 112.9~\kms. Abundance relative to \HM: $2\times10^{-6}$ \\
$^c$~blue wing from 88.9 to 91.9~\kms, red wing from 100.9 to 103.9~\kms. Abundance relative to \HM: $10^{-8}$ \\
$^d$~blue wing from 90.1 to 92.5~\kms, red wing from 99.1 to 101.5~\kms. Abundance relative to \HM: $2\times10^{-10}$
\end{flushleft}
\end{table}

From Table~\ref{tout} one notes that the values
increase with decreasing abundance of the molecule. This can be explained
in terms of decreasing optical depth, because higher opacities lead to
an underestimate of the column density and hence of the mass. Indeed,
from the ratio between the \CO\ and \COI\ emission in the line wings
one derives opacities as high as $\sim$30 for \CO.
Therefore,
the most reliable estimates should be those obtained from \MCNII. These
depend significantly on the \MCN\ abundance, which is known to present
considerable variations in molecular clouds. However, the value assumed in our
calculations is one of the highest found in the literature and we therefore
believe that our estimates are likely to be lower limits.
This leads us to another consideration. Our estimates are very close to the
largest outflow parameters ever measured (see e.g.
L\'opez-Sepulcre et al.~\cite{lopsep} and Wu et al.~\cite{wu04}),
corresponding to a YSO powering the outflow of at least
$\sim10^5~L_\odot$. More precisely, using Wu et al.~(\cite{wu04}) relationship
between bolometric luminosity ($L_{\rm bol}$) of the powering source and
outflow momentum rate, from the value in Table~\ref{tout} one obtains
$L_{\rm bol}=6\times10^6$~\Lsun. This is much higher than the luminosity
estimated for the HMC ($\sim10^5~L_\odot$) by Osorio et al.~(\cite{osor}) and
that obtained from the corresponding IRAS fluxes ($2.6\times10^5$~\Lsun; see
Cesaroni et al.~\cite{cesa94}). Moreover, the momentum rate of
0.3~\Msun\,\kms\,yr$^{-1}$ is an order of magnitude higher than that
estimated by Cesaroni et al.~(\cite{cesa10}) from VLA observations of the
free-free continuum emission ($\sim$0.03--0.06~\Msun\,\kms\,yr$^{-1}$).

It is also worth noting that the velocity gradient in G31.41 seems to involve
the whole core and not only the gas emitting in the line wings. Usually, in
outflow sources the emission close to the systemic velocity -- i.e. that
inside the FWHM of the line -- traces the molecular core; in contrast, in
G31.41 most of the \MCN\ emission is affected by the velocity gradient, as
one can see from the channel maps in Fig.~\ref{fmcnchm}, where only emission
over a limited velocity range ($\sim$2--3~\kms) appears to arise from the
central region. As discussed below in Sect.~\ref{sotto}, this situation
is reminiscent of that observed in circumstellar disks around lower-mass YSOs. Our
estimates of the outflow parameters based only on the line-wing emission are
likely lower limits, which makes the case of
G31.41 even more extreme compared to typical outflows.

In conclusion, the outflow hypothesis seems to yield values for the
outflow parameters that are too high.
In addition, one should note that the typical parameters
given in studies such as those quoted above refer to single-dish
observations of pc-scale flows, an order of magnitude higher than that
observed in G31.41. Although we cannot exclude that in G31.41 one is
observing the earliest stages of the expansion, it is questionable that the
parameters of a young, compact outflow are greater than those typical of much
older, extended outflows.

Finally, we note that the dynamical time scale of the putative outflow is an
order of magnitude shorter than the time needed to form typical hot core species
-- such as methyl cyanide -- according to theoretical models (see e.g.
Charnley et al.~\cite{charn}). Our estimate of $t_{\rm dyn}$ is
affected by large uncertainties, owing to the unknown inclination angle and the
difficulty in tracing the whole extent of the lobes. However, to increase
$t_{\rm dyn}$ by a factor $\ga$10 one has to assume that the outflow lies
very close ($\la$6\degr) to the line of sight and/or that we detect only the
most compact part of lobes extending over a region 10 times larger than that
imaged in our \MCN\ and CO maps.
The latter explanation is ruled out by our SMA+30m combined maps that do not
reveal any large-scale bipolar outflow. The former would imply a significant
overlap between the blue- and red-shifted lobes in the plane of the sky,
which is not seen in Fig.~\ref{fhmc}. Indeed, from the observed separation
between the peaks of the red- and blue-shifted emission ($\sim$1\arcsec),
assuming that the intrinsic length of the \MCN\ lobes is comparable to the
radius of the \MCN\ core measured in the plane of the sky ($\sim$1\arcsec),
one can obtain a rough estimate of the inclination angle of
$\arcsin(0.5)=30\degr$. Correspondingly, $t_{\rm dyn}$ would increase by only
a factor 1.7.

\begin{figure*}
\centering
\resizebox{13cm}{!}{\includegraphics[angle=0]{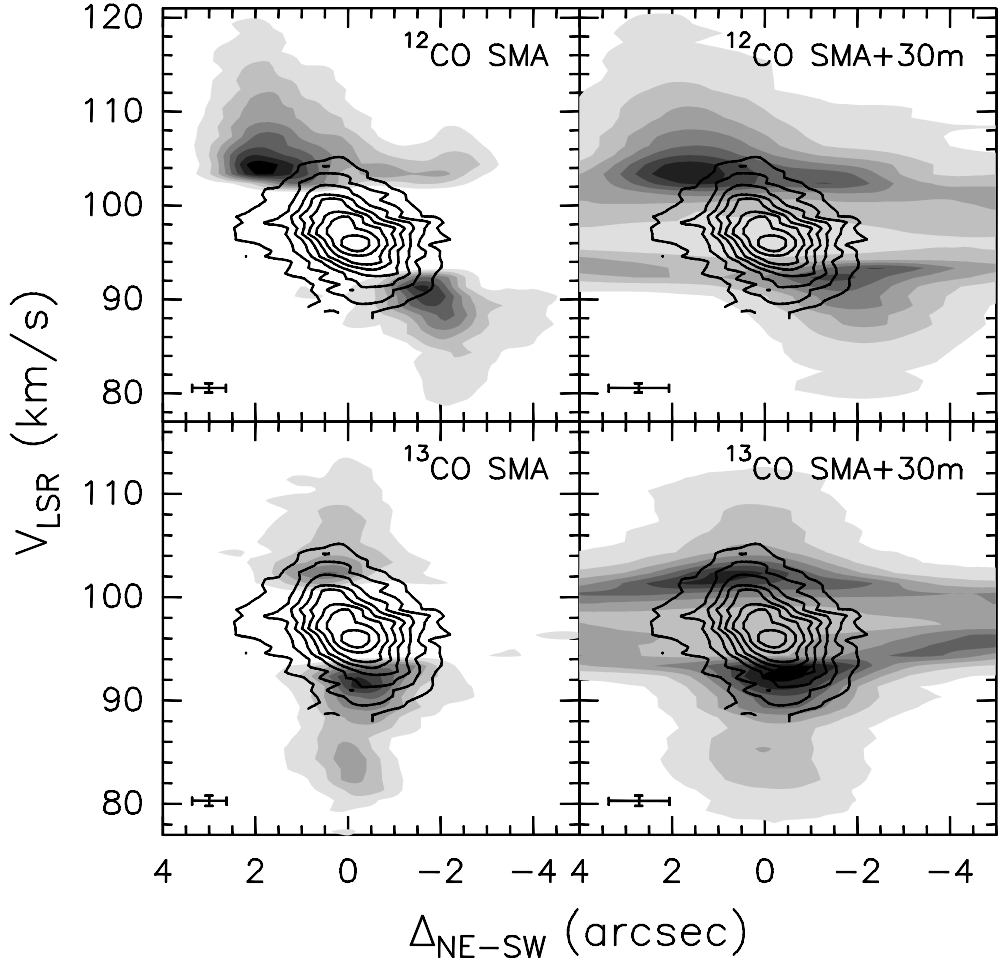}}
\caption{
Position-velocity plots along the direction with P.A.=68\degr\ passing through
the HMC. The offset is measured from the phase center, positive toward NE.
Contours correspond the \MCN(12--11) $K$=4 component emission, while
gray scales indicate the \CO(2--1) and \COI(2--1) line emission, as indicated
in each box. Contour levels increase from 0.2 in steps of 0.2~Jy/beam for
\MCN, from 0.24 in steps of 0.24~Jy/beam for \CO\ SMA, from 0.6 in steps of
1~Jy/beam for \CO\ SMA+30m, from 0.12 in steps of 0.175~Jy/beam for \COI\ SMA,
and from 0.15 in steps of 0.5~Jy/beam for \COI\ SMA+30m.
The cross in the bottom left of each box denotes the angular and spectral
resolutions.
}
\label{fpvmcnco}
\end{figure*}

Given the number of problems encountered in the outflow scenario,
we investigate also a second possibility,
namely that the velocity gradient of the HMC is due to a rotating toroid.

\subsubsection{Rotating toroid}
\label{storo}

In an attempt to shed light on the velocity field in the HMC and its
molecular surroundings, we overlay in Fig.~\ref{fpvmcnco} the position-velocity
plots along the velocity gradient in high- and low-density tracers. 
For the sake of completeness we show both the SMA and the combined 30m+SMA
data in the \CO\ and \COI\ (2--1) transitions. Note that the cut was made
along P.A.=68\degr\ and all plots were obtained after averaging the
emission along the direction perpendicular to the cut. This has the twofold
purpose of increasing the S/N of the plots and taking into account emission
at all offsets along the (putative) rotation axis.

As already explained, the \CO\ and \COI\ emission close to the systemic
velocity is highly opaque and does not convey any information on the HMC, but
is instructive to study the high-velocity and/or low-density gas around it.
Vice versa, the \MCN\ emission is an excellent HMC tracer, but is not
detected at high velocities and/or beyond $\sim$2\arcsec\ from the center.
Combining all these tracers is the only way to perform a detailed and
complete analysis of the velocity field in this core. Indeed,
Fig.~\ref{fpvmcnco} is very instructive. It shows that the velocity trend
observed in \MCN\ is complementary to that seen in the CO isotopomers
and fills the gap caused by the interferometer resolving out the extended, bulk CO
emission.

A more thorough inspection of this figure reveals that, despite the similarity, the
\CO\ and \COI\ plots are significantly different. While the \COI\ and \MCN\
patterns match very well in the overlapping regions (i.e. between
89 and 94~\kms\ and between 102 and 105~\kms), the \CO\ emission is
offset by $\sim$1\arcsec\ from the \MCN\ and \COI\ patterns, to the SW
in the blue-shifted part, and to the NE in the red-shifted part.
An
interesting difference between \MCN\ and \COI, on the one hand, and
\CO, on the other, is that the high-velocity emission is found at relatively
large offsets in \CO, whereas in \COI\ it appears to peak close to the
HMC center.

Is it possible to find a coherent interpretation of all these facts in the
rotating toroid scenario? A possible explanation is that of a
self-gravitating structure with a heavy, compact stellar cluster at the
center. In the outer regions (traced by \CO), the gravitational field is
determined by the gas mass and the rotation curve flattens, whereas close to
the center (where \COI\ and \MCN\ are detected) the velocity field resembles
Keplerian rotation, because the gas mass becomes comparable to the stellar
mass. This scenario is analogous to that studied by Bertin \& Lodato
(\cite{belo}), who determined the rotation curve of a self-gravitating disk
with a central star.

In order to check the viability of this interpretation, below we
first estimate the mass of the HMC and then verify whether this suffices to
support rotational equilibrium.

From the continuum flux (see Table~\ref{tcont}), we derive a mass of
1700~$M_\odot$ for the HMC, assuming a dust temperature of 100~K and a dust
absorption coefficient at 1.3~mm of 0.005~cm$^2$~g$^{-1}$ (Kramer et
al.~\cite{kramer}). The temperature of 100~K is a lower limit, because the real
temperature must be higher than or equal to the maximum brightness
temperature of the \MCN\ lines measured in the synthesized beam ($\sim$95~K).
The latter corresponds to the surface temperature of the HMC, because the
ground state \MCN\ lines are optically thick and, as discussed by BEL05, the
temperature inside the HMC is likely increasing.
The mass obtained is affected by a large uncertainty mostly because of the
poorly known dust properties: for example, BEL04 adopted an absorption
coefficient of 0.02~cm$^2$~g$^{-1}$, which would imply a HMC mass of only
420~$M_\odot$.

\begin{figure}
\centering
\resizebox{8.5cm}{!}{\includegraphics[angle=-90]{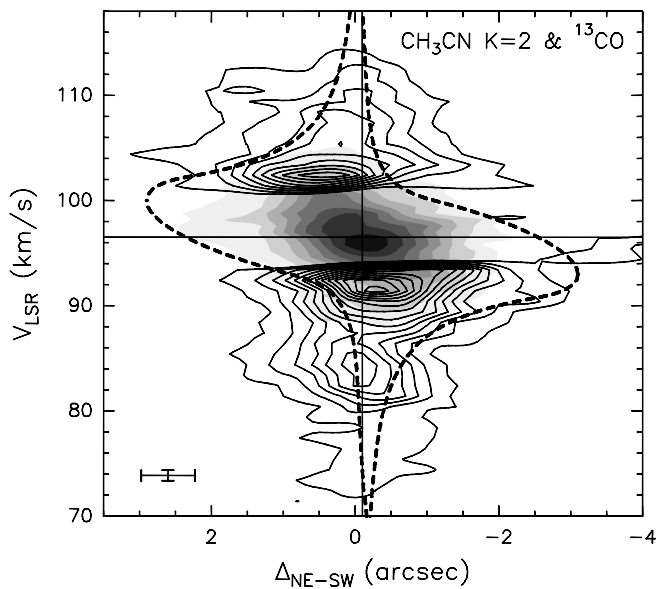}}
\caption{
Overlay of the \COI(2--1) (only SMA data; contours) and \MCN(12--11)~$K$=4
position-velocity plots along the direction with P.A.=68\degr\ passing
through the HMC. The offset is measured from the phase center, positive
toward NE. Contour levels increase from 0.2 in steps of 0.2~Jy/beam for
\MCN\ and from 0.1 in steps of 0.1~Jy/beam for \CO. The vertical and
horizontal solid lines denote the position of the continuum
peak and the systemic velocity, respectively. The thick dashed pattern encompasses the
region inside which emission is expected from a disk with radius of
3\arcsec\ (or 0.11~pc) undergoing free-fall and Keplerian rotation about
330~\Msun. The cross in the bottom left denotes the angular and spectral
resolutions.
}
\label{fpvfit}
\end{figure}

In 
Fig.~\ref{fpvfit} we show the same position-velocity plot of the \MCN\ and
\COI\ lines as in Fig.~\ref{fpvmcnco}, this time overlaying the pattern
outlining the region inside which emission is expected for a Keplerian
rotating and free-falling disk. The latter does not take into account the
line-width nor the spectral and angular resolutions and has been obtained
assuming that the gas velocity is the vector sum of a tangential component due
to Keplerian rotation about a central mass, $M$, plus a radial component due
to free-fall onto the same mass.
Under these assumptions the velocity component along the line of sight can be
expressed as
\begin{equation}
 V = \sqrt{G\,M}\frac{x}{R^\frac{3}{2}} + \sqrt{2G\,M}\frac{z}{R^\frac{3}{2}}~,
\end{equation}
where $x$ and $z$ are the coordinates, respectively, along the disk plane and
the line of sight, and $R=\sqrt{x^2+z^2}$ is the distance from the center of
the disk. We also assume that R lies between the disk radius,
$R_{\rm o}$, and a minimum inner radius, $R_{\rm i}$.  The first term on the
right hand side of the equation is the component due to Keplerian rotation,
the second that due to free fall. The dashed pattern in Fig.~\ref{fpvmcnco}
has been obtained by plotting the maximum and minimum velocities $V$ for $z$
varying across the disk, i.e. from $-\sqrt{R_{\rm o}^2-x^2}$ and
$+\sqrt{R_{\rm o}^2-x^2}$, taking into account that the region $R<R_{\rm i}$
is forbidden.
In our case,
a satisfactory fit is obtained for
$M=330~M_\odot$, $R_{\rm o}=3\arcsec$ (or
0.11~pc), and $R_{\rm i}=0\farcs15$ (or 0.0055~pc).
Note that the signature of (pseudo-)Keplerian rotation is the
``butterfly'' shape of the plot, determined by the two ``spurs'' of emission
at
about $\pm$3\arcsec\ and $\pm$4~\kms\
relative to the systemic velocity
($\sim$96.5~\kms), plus the presence of high-velocity emission at zero
offset\footnote{
The peak in the contour plot at 0\arcsec\ and $\sim$84~\kms\ is
likely due to an unidentified line of a core tracer,
whereas the broad wing of emission extending up to
$\sim$70~\kms\ is genuine high-velocity \COI(2--1) emission.
}. It is worth stressing that the contribution of the \COI(2--1) line is
crucial to outline such a pattern, which cannot be recognized from the sole
\MCN\ emission. This explains why BEL05 managed to fit
their \MCN\ data assuming a flat or solid-body rotation curve,
and Girart et al.~(\cite{gira}), from a number of CH$_3$OH lines, found that
the rotation velocity increases with radius.

One may wonder whether a HMC as massive as several 100~$M_\odot$ may be
undergoing Keplerian rotation. As suggested by Cesaroni et al.~(\cite{ppv}) and
demonstrated by Beltr\'an et al.~(\cite{bel11}), these massive, large
rotating cores are to be considered transient toroidal structures feeding
a cluster of YSOs rather than stable circumstellar accretion disks. In fact,
the latter are stabilized by the central star(s), whose mass is greater
than that of the disk, whereas the former are dynamically dominated by the
gas mass and hence are short-lived. Although this scenario may be true in most
cases, G31.41 might represent an exception. We speculate that this HMC could
contain a large number of stars tightly packed in the central region. In this
case the stars could have a stabilizing effect analogous to that of a single
point-like object located at the HMC center, similar to the previously
mentioned model by Bertin \& Lodato~(\cite{belo}).

With all this in mind, the pseudo-Keplerian pattern recognized in
Fig.~\ref{fpvfit} suggests that the HMC mass should be comparable to the
total mass of the embedded stars that are tightly packed at the center. Is 
this scenario plausible? Indeed, in the case of G10.62--0.38, a
massive star forming region with a luminosity of $9.2\times10^5~L_\odot$,
Sollins et al. (\cite{sollins}) suggest that within a radius of 0.03~pc,
several O stars with a total mass of 175~\Msun\ have formed at the center of
a flattened disk. In our case, the
existence of (at least) two high-mass YSOs close to the HMC
center and separated (in projection) by only 0\farcs19 or 1480~AU has been
proved by Cesaroni et al.~(\cite{cesa10}), who detected the free-free
continuum emission. In terms of mass, these two YSOs represent only the tip of
the iceberg, because the total mass of the associated cluster is dominated by
(undetected) low-mass stars. Assuming e.g. a Miller \& Scalo (\cite{milsc})
mass function,
one can estimate the cluster mass from that of the most massive star. For
this we can use the value of 20--25~\Msun\ computed by Osorio et
al.~(\cite{osor}) from their model fit and obtain a total mass of the cluster
of $\sim10^3$~\Msun.
The same result is obtained by fixing the total luminosity of the cluster
to the value of $2\times10^5~\Lsun$ quoted by Osorio et al.~(\cite{osor}).
In all likelihood $\sim10^3$~\Msun\ is
an upper limit to the
real stellar mass, because the cluster mass function may be highly incomplete
in such a small region. However, this estimate indicates that
the dynamical mass of $\sim$330~\Msun, obtained from the fit in
Fig.~\ref{fpvfit}, is probably dominated by the stellar mass,
which would lend
support to the application of Bertin \& Lodato~(\cite{belo}) model to the
case of G31.41.

Finally, we note that the toroid interpretation is compatible with the
hourglass-shaped morphology of the magnetic
field (see Girart et al.~\cite{gira}), whose symmetry axis is directed SE--NW,
because the latter coincides with the rotation axis of the toroid.
However,
a caveat is in order. Girart et al.~(\cite{gira}) find a significant
correlation between rotation velocity and radius in the core, using different
tracers. In particular, they conclude that in the region sampled by their
observations (lying between $\sim$0\farcs5 and 1\farcs6), velocity
increases with distance from the HMC center (see their Fig.~4). This result
seems inconsistent with our hypothesis that the core is undergoing
pseudo-Keplerian rotation, because in this case the rotation velocity should
decrease with radius. Perhaps this discrepancy can be explained by the
presence of the magnetic field.
Higher angular resolution
observations are needed to sample the velocity field inside the core
and thus obtain a reliable, direct measurement of the rotation curve.

\subsection{The \MCN\ ``8-shaped'' structure}
\label{sotto}

The last aspect we will discuss in the context of the outflow/toroid
controversy, is the double-peaked, ``8-shaped'' structure
observed in the \MCN\ $K$=2 map at the systemic velocity (see
Fig.~\ref{fmapsd}). The same morphology is seen in all components up to $K$=6,
i.e. up to excitation energies of $\sim$300~K, whereas the $K$=8 map (energy
of 513~K) presents a single peak at the HMC center (Fig.~\ref{fmapsd}).
In Sect.~\ref{score} we interpreted these facts in terms of opacity
and temperature gradients. Indeed, BEL05 found
that the \MCN\ rotational temperature peaks toward the HMC center and
from the ratio between our \MCN\ and \MCNII\ (12--11) $K$=2 data we calculate
an optical depth at the peak of the line in the range 8--77 across the HMC.
The interplay between opacity and temperature can explain the
existence of a dip at the HMC center, but does not justify the lack of
circular symmetry in the low-$K$ maps. Here, we wish to find an explanation
for the ``8-shaped'' feature and check whether this can better fit into
the outflow or toroid model.

\begin{figure}
\centering
\resizebox{8.5cm}{!}{\includegraphics[angle=-90]{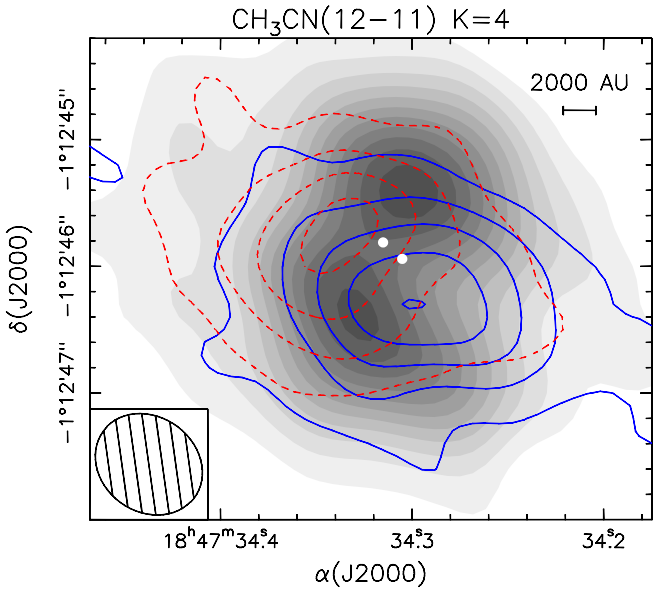}}
\caption{
Maps of the blue- (solid contours) and red-shifted (dashed) emission in the
\MCN(12--11) $K$=4 line overlayed on the bulk emission (gray scale) in the
same line. The integration was performed over the
velocity ranges 88.9--91.3~\kms, 100.9--103.3~\kms, and 95.2--96.7~\kms, respectively.
Contour levels increase from 0.15 in steps of 0.27~Jy/beam for the blue- and
red-shifted gas, and from 0.2 in steps of 0.2~Jy/beam for the bulk emission.
The synthesized beam of the images is drawn in the bottom left. The two white
points mark the positions of the two compact free-free sources detected
by Cesaroni et al.~(\cite{cesa10}).
}
\label{fhmc}
\end{figure}

As a basis for our discussion, in Fig.~\ref{fhmc} we present an overlay of
the blue- and red-shifted \MCN\ $K$=4 emission on the bulk emission in
the same line. In this way, one is comparing the high- with the low-velocity
emission.

In the outflow scenario, a na\"{\i}ve interpretation of this figure is that
the high-velocity gas is leaking from the HMC through the axis of a
``donut-like'' structure seen edge-on, corresponding to the
``8-shaped'' feature in the map. In this case, the two peaks would coincide
with the
maxima of column density across the ``donut''. Albeit plausible, this
interpretation has a problem. The large \MCN\ opacity should prevent the
detection of two distinct peaks, because the line brightness is independent of
column density. Therefore, instead of the 8-shaped feature, one should see an
elongated structure perpendicular to the bipolar outflow and peaking {\it at
the center} of it (i.e. at the HMC center).

Explaining the presence of the 8-shaped structure appears to be a problem
also in the toroid scenario. In fact, if the toroid is seen edge-on, the
emission at the systemic velocity should peak in between the blue- and
red-shifted emission. However, if the toroid is inclined with respect to the
line of sight, the nearest and farthest sides of it are seen displaced from
one another on the plane of the sky, symmetrically displaced with respect to
the center. This is indeed what we see in Fig.~\ref{fhmc}, where the two
peaks lie along the (projection of the) rotation axis and are equally offset
from the center. The situation depicted in this figure resembles that of the
ring undergoing Keplerian rotation about the low-mass system GG~Tau (see
Fig.~6a of Guilloteau et al.~\cite{ggtau}). Clearly, this comparison is
inappropriate, because the scales of the two objects differ by more than an
order of magnitude. Moreover, G31.41 is probably more similar to a
``puffy pancake'' rather than a ``donut'' or ring, because
gas and dust are present at all
radii, as demonstrated by the continuum and high-energy line emission peaking
at the center of the HMC (see Figs.~\ref{fmapsu} and~\ref{fmapsd}). Note,
however, that the emission from the lower energy lines may be arising mostly
from the colder outer region and hence be confined in a ring. Finally,
G31.41, unlike GG~Tau, is embedded in a dense pc-scale envelope, which
complicates the line and continuum radiative transfer. Despite all these
differences, the comparison between the two objects is intriguing and may
qualitatively explain the features observed in G31.41.

Further support for this interpretation is obtained by comparing the case of
G31.41 to that of a source closer in mass than GG~Tau. This is the
Herbig Ae star MWC\,758, which is surrounded by a circumstellar disk, as
demonstrated by the model fit of Isella et al.~(\cite{isella}). A comparison
between the channel maps presented by these authors (see their Fig.~2) with
those in Fig.~\ref{fmcnchm} reveals a surprising similarity: in both
cases the shape of the emission is circular and peaks at opposite positions
with respect to the center in the blue- and red-shifted channels, but turns
into an 8-shaped structure close to the systemic velocity. This fact may be
the fingerprint of a rotating disk in G31.41, as well as in MWC\,758.

\section{Summary and conclusions}
\label{scon}

With the present study we have performed a deep analysis of the geometrical
structure and kinematics of the HMC in the high-mass star-forming region G31.41.
Thanks to new high angular resolution SMA images and complementary IRAM 30-m
maps, we have investigated both the HMC, through the \MCN(12--11) transition,
and the lower density surroundings, through \CO\ and \COI. We
could thus search for a possible outflow powered by the star(s)
embedded in the HMC. In particular, our main goal was to shed light on the
nature of the NE--SW velocity gradient, previously detected in the HMC
by various authors, and possibly distinguish between rotation and expansion.

Our new data confirm the presence of the velocity gradient basically in all
HMC tracers observed, and indicate that a red--blue symmetry along the same
axis is seen also in the CO isotpomers. We were unable to detect any
bipolar outflow along the SE--NW direction, in contrast with the findings
of Olmi et al.~(\cite{olmi96b}). The latter was probably an artifact caused
by very limited uv sampling and lack of zero-spacing information.

We find a hint of infall in the
HMC, consistent with the results of Girart et al.~(\cite{gira}). A comparison
between the \MCN\ and CO maps suggests that these species are tracing the
same phenomenon. Especially the \COI(2--1) line emission appears to be the
obvious prosecution of the \MCN(12--11) emission on a larger scale.

We have discussed whether the velocity gradient observed in G31.41 is a
compact bipolar outflow or a rotating, geometrically thick, toroidal
structure. The former hypothesis implies outflow parameters typical of
high-mass stars with luminosities in excess of $10^5$~\Lsun, whereas previous
estimates by Osorio et al.~(\cite{osor}) and Cesaroni et
al.~(\cite{cesa94,cesa10}) indicate a significantly lower luminosity for
G31.41. Moreover, such an outflow would be much more compact than typical
bipolar outflows detected in high-mass star-forming regions and the dynamical
time scale would be an order of magnitude shorter than that needed to form HMC
spacies such as \MCN. It is also worth noting that a core undergoing
expansion seems difficult to reconcile with the detection of infall.

The composite \MCN\ and \COI\
position-velocity plot can be explained if the gas is undergoing both
(pseudo-)Keplerian rotation and infall, suggesting that this massive object
could be dynamically stabilized by a compact cluster of YSOs tightly packed
inside a few 1000~AU from the center. In this scenario the \CO\ emission
could be tracing the outer regions, where the gravitational field is dominated
by the gas and the rotation velocity tends to a constant value.
Albeit intriguing, this interpretation remains speculative and could be
proved only by producing synthetic \MCN\ and CO maps and position-velocity
plots to be compared with the data. This numerical effort goes beyond the
purposes of the present study.

We conclude that the case of G31.41 could be a scaled-up version of lower-mass
YSOs, such as GG~Tau and MWC\,758, where circumstellar disks have been
detected and successfully modeled. We stress that whatever the
interpretation, the observations of this object have produced an important
finding for the study of high-mass star formation. We obtained iron-clad
evidence for the existence of a velocity gradient across the HMC and
demonstrated
that this gradient is {\it not} due to multiple cores unresolved in the
beam and moving at different velocities, but to the gas undergoing a smooth
drift in (line-of-sight) velocity from the one end to the other of the core.
If, as we believe, one is dealing with a rotating toroid, then this indicates
that star formation in this HMC may proceed in a similar way to low-mass
stars, namely through circumstellar, centrifugally supported disks ``hidden''
inside the densest interiors of the toroid itself. If, instead, the outflow
interpretation were correct, such a sharp, neat velocity gradient would
strongly suggest the existence of a very effective focusing mechanism for the
outflow and the best candidate as a collimating agent would be a (yet
undetected) circumstellar disk. Therefore, regardless of whether the G31.41
HMC is undergoing expansion or rotation, the role of disks in the mechanism
of high-mass star formation seems to be strengthened by our findings. The
advent of ALMA will be crucial to detect deeply embedded disks in HMCs.

\begin{acknowledgements}
We thank the SMA staff for their help during the observations. The GILDAS team
is also acknowledged for the excellent software that we used to analyze our data.
\end{acknowledgements}


\begin{thebibliography}{}

\bibitem[2008]{araya}
 Araya, E., Hofner, P., Kurtz, S., Olmi, L., \& Linz, H. 2008, ApJ, 675, 420
\bibitem[2004]{bel04}
 Beltr\'an, M.T, Cesaroni, R., Neri, R., et al. 2004, ApJ, 601, L187 (BEL04)
\bibitem[2005]{bel05}
 Beltr\'an, M.T, Cesaroni, R., Neri, R., et al. 2005, A\&A, 435, 901 (BEL05)
\bibitem[2011]{bel11}
 Beltr\'an, M.T, Cesaroni, R., Neri, R., \& Codella, C. 2011, A\&A, 525, A151
\bibitem[2003]{benj}
 Benjamin, R.A., Churchwell, E., Babler, B.L., et al. 2003, PASP, 115, 953
\bibitem[1999]{belo}
 Bertin, G., Lodato, G. 1999, A\&A 350, 694
\bibitem[1994a]{cesa94}
 Cesaroni, R., Churchwell, E., Hofner, P., Walmsley, C.M., \& Kurtz, S. 1994a,
 A\&A, 288, 903
\bibitem[1994b]{cesag31}
 Cesaroni, R., Olmi, L., Walmsley, C.M., Churchwell, E., \& Hofner, P. 1994b,
 ApJ, 435, L137
\bibitem[1998]{cesa98}
 Cesaroni, R., Hofner, P., Walmsley, C.M., \& Churchwell, E. 1998, A\&A,
 331, 709
\bibitem[2007]{ppv}
 Cesaroni, R., Galli, D., Lodato, G., Walmsley, C.M., \& Zhang, Q. 2007, in
 Protostars and Planets V, ed. B. Reipurth, D. Jewitt, \& K. Keil
 (Tucson: Univ. of Arizona Press), 197
\bibitem[2010]{cesa10}
 Cesaroni, R., Hofner, P., Araya, E., \& Kurtz, S. 2010, A\&A, 509, 50
\bibitem[1992]{charn}
 Charnley, S.B., Tielens, A.G.G.M., \& Millar, T.J. 1992, ApJ, 399, L71
\bibitem[1987]{gamu}
 Gaume, R.A. \& Mutel, R.L. 1987, ApJS, 65, 193
\bibitem[2004]{gibb}
 Gibb, A.G., Wyrowski, F., \& Mundy, L.G. 2004, ApJ, 616, 301
\bibitem[2009]{gira}
 Girart, J.M., Beltr\'an, M.T., Zhang, Q., Rao, R., \& Estalella, R. 2009,
 Science, 324, 1408
\bibitem[1999]{ggtau}
 Guilloteau, S., Dutrey, A., \& Simon, M. 1999, A\&A, 348, 570
\bibitem[1996]{ho2004}
 Ho, P.T.P., Moran, J.M., \& Lo, K.Y. 2004, ApJ, 616, L1
\bibitem[2010]{isella}
 Isella, A., Natta, A., Wilner, D., Carpenter, J.M., \& Testi, L. 2010, ApJ, 725, 1735
\bibitem[2007]{keto07}
 Keto, E.H. 2007, ApJ, 666, 976
\bibitem[2003]{kramer}
 Kramer, C., Richer, J., Mookerjea, B., Alves, J., Lada, C. 2003, A\&A 399, 1073
\bibitem[2009]{krum}
 Krumholz, M.R., Klein, R.I., McKee, C.F., Offner, S.S.R., \& Cunningham, A.J. 2009, Science, 323, 754
\bibitem[2010]{kuip}
 Kuiper, R., Klahr, H., Beuther, H., \& Henning, Th. 2010, ApJ, 722, 1556
\bibitem[2009]{lopsep}
 L\'opez-Sepulcre, A., Codella, C., Cesaroni, R., Marcelino, N.,
 \& Walmsley, C.M. 2009, A\&A, 499, 811
\bibitem[1979]{milsc}
 Miller, G.E. \& Scalo, J.M. 1979, ApJ, 41, 513
\bibitem[1987]{nakano}
 Nakano, T. 1987, MNRAS, 224, 107
\bibitem[1996]{olmi96b}
 Olmi, L., Cesaroni, R., Neri, R., \& Walmsley, C.M. 1996, A\&A 315, 565
\bibitem[2009]{osor}
 Osorio, M., Anglada, G., Lizano, S., \& D'Alessio, P. 2009, ApJ 694, 29
\bibitem[1993]{past}
 Palla, F., Stahler, S.W. 1993, ApJ 418, 414
\bibitem[1986]{sand}
 Sanders, D.B., Clemens, D.P., Scoville, N.Z., \& Solomon, P.M. 1986, ApJS, 60, 1
\bibitem[2004]{sault}
 Sault, R.J., Teuben, P.J., \& Wright, M.C.H. 1995, in ASP Conf. Ser. 77, Astronomical Data Analysis Software and Systems IV, 433
\bibitem[2001]{simon}
 Simon, M., Guilloteau, S., Dutrey, A. 2001, ApJ 545, 1034
\bibitem[2005]{sollins}
 Sollins, P.K., Zhang, Q., Keto, E., \& Ho, P.T.P. 2005, ApJ 624, L49
\bibitem[1993]{vand}
 Van Dishoeck, E.F., Blake, G.A., Draine, B.T., \& Lunine, J.I. 1993,
 Protostars and Planets III, ed. E.H. Levy \& J.I. Lunine,
 (Tucson: Univ. of Arizona Press), 163
\bibitem[1994]{wiro}
 Wilson, T.L. \& Rood, R. 1994, ARA\&A, 32, 191
\bibitem[2004]{wu04}
 Wu, Y., Wei, Y., \& Zhao, M., 2004, A\&A, 426, 503

\end{thebibliography}
\end{document}